\documentclass[11pt,a4paper,american]{extarticle}
\usepackage[T1]{fontenc}
\usepackage[latin1]{inputenc}
\usepackage{amsmath}
\usepackage{amssymb}
\usepackage{setspace}
\usepackage{esint}
\usepackage[numbers]{natbib}
\makeatletter

\usepackage{babel}
\usepackage{bbm}
\usepackage{hyperref}
\usepackage{authblk}
\usepackage{tikz}
\usetikzlibrary{matrix}
\usepackage{cancel}
\date{}
\allowdisplaybreaks
\numberwithin{equation}{section}

\author{Renann Lipinski Jusinskas\thanks{renannlj@fzu.cz}}
\affil{Institute of Physics of the Czech Academy of Sciences \\ CEICO - Central European Institute for Cosmology and Fundamental Physics
\authorcr  Na Slovance 2, 182 21, Prague - Czech Republic}

\makeatother

\usepackage{babel}
\begin{document}
\title{Towards the underlying gauge theory of the pure spinor superstring}
\maketitle
\begin{abstract}
Previous attempts to determine the worldsheet origin of the pure spinor
formalism were not completely successful, but introduced important
concepts that seem to be connected to its fundamental structure, \emph{e.g.},
emergent supersymmetry and the role of reparametrization symmetry.

In this work, a new proposal towards the underlying gauge theory of the
pure spinor superstring is presented, based on an extension of Berkovits'
twistor-like constraint. The gauge algebra is analyzed in detail and
worldsheet reparametrization is shown to be a redundant symmetry.
The master action is built with a careful account of the intrinsic
gauge symmetries associated with the pure spinor constraint and a
consistent gauge fixing is performed. After a field redefinition,
spacetime supersymmetry emerges and the resulting action describes
the pure spinor superstring.\tableofcontents{}
\end{abstract}

\section{Overview\label{sec:overview}}

The pure spinor formalism of the superstring was introduced by Berkovits
almost two decades ago \cite{Berkovits:2000fe}. Since then, it has
been studied and explored in many different aspects, taking advantage
of its symmetry preserving character and bosonic string-like amplitude
prescription. These aspects range from the impressive $3$-loop computation
of scattering amplitudes of \cite{Gomez:2013sla} or the recent $N$-point
$1$-loop results of \cite{Mafra-Schlotterer}, to the investigation
of the quantization of the superstring in the $AdS_{5}\times S^{5}$
background (see \cite{Mazzucato:2011jt} for a review and references
therein) or to the analysis of supersymmetry breaking effects in the
superstring \cite{Berkovits:2014rpa}.

There is abundant evidence that the pure spinor superstring is related
to the spinning string \cite{Ramond:1971gb,Neveu:1971rx} and to the
Green-Schwarz superstring \cite{Green:1980zg,Green:1981yb}. Scattering
amplitudes computed in the pure spinor superstring were shown to be
equivalent to the spinning string amplitudes up to two loops \cite{Berkovits:2005ng}.
Furthermore, it has been shown in \cite{Berkovits:2000nn} that the
pure spinor cohomology in the light-cone gauge describes the usual
physical spectrum of the superstring. Later on this equivalence was
explored in \cite{Berkovits:2004tw} and more recently in \cite{Berkovits:2014bra},
where a combination of field redefinitions and similarity transformations
helped to identify the Green-Schwarz and the pure spinor superstrings.
In \cite{Jusinskas:2014vqa}, the DDF-like structure of the pure spinor
cohomology was finally made explicit. From another perspective, superembedding
techniques (see \cite{Sorokin:1999jx} for a review and references
therein) seem to provide a fertile ground for exploring the classical
equivalence between the different superstring formalisms (\emph{e.g.}
\cite{Berkovits:1989zq,Tonin:1991ii}). The superembedding origin
of the pure spinor description of the heterotic superstring was demonstrated
in \cite{Matone:2002ft}.

However, the pure spinor formalism lacks a fundamental worldsheet
description, meaning (1) a two-dimensional reparametrization invariant
gauge theory which upon quantization concretely leads to its characteristic
BRST structure and (2) a first principles derivation of the pure spinor constraint itself. 
The goal of this work is to present a possible resolution for the point (1), but still assuming that the pure spinor is a fundamental variable.

\subsection*{The twistor-like constraint}

The first step to understand the gauge structure of the pure spinor
formalism from a more fundamental point of view was taken in \cite{Berkovits:2011gh}
with the introduction of the twistor-like constraint
\begin{eqnarray}
H_{\alpha} & \equiv & P_{m}(\gamma^{m}\lambda)_{\alpha},\nonumber \\
 & = & 0,\label{eq:TLconstraint}
\end{eqnarray}
where $P_{m}$ is the canonical conjugate of the target-space coordinate
$X^{m}$, with $m=0,\ldots,9$, $\gamma_{\alpha\beta}^{m}$ denotes
the chiral blocks of the Dirac matrices, with $\alpha=1,\ldots,16$,
and $\lambda^{\alpha}$ is a pure spinor variable satisfying
\begin{equation}
(\lambda\gamma^{m}\lambda)=0.
\end{equation}
The novel feature of this approach was that $\lambda^{\alpha}$ appeared
as a fundamental variable in the worldline/worldsheet\footnote{Twistor-like variables arise naturally using superembedding techniques
and the first appearance of pure spinors in this context was in \cite{Tonin:1991ii}.} and the superpartners of $X^{m}$, denoted by $\theta^{\alpha}$,
entered the formalism as ghost fields associated to the gauge symmetry
generated by \eqref{eq:TLconstraint}. In this model, supersymmetry
is an emergent feature related to a ghost twisting operation on the
gauge fixed action. However, the gauge symmetries related to the pure
spinor constraint were not completely considered in this approach,
leading to an incorrect description of the ghost fields.

A new attempt to quantize the twistor-like constraint was made in
\cite{Berkovits:2014aia}, with a different mechanism for the emergence
of spacetime supersymmetry. The problem of this proposal was an overconstrained
action which ultimately leads to a trivialization of the model.

This flaw was later corrected in \cite{Berkovits:2015yra}, where
a new gauge theory was proposed to explain the origin of the pure
spinor formalism. Berkovits' first order action can be cast as
\begin{multline}
S_{B}=\int d\tau d\sigma\{P_{m}\partial_{\tau}X^{m}+w_{\alpha}\nabla_{\tau}\lambda^{\alpha}+\hat{w}_{\hat{\alpha}}\hat{\nabla}_{\tau}\hat{\lambda}^{\hat{\alpha}}+K_{\alpha}\nabla_{\sigma}\lambda^{\alpha}+\hat{K}_{\hat{\alpha}}\hat{\nabla}_{\sigma}\hat{\lambda}^{\hat{\alpha}}\\
-\tfrac{1}{2}L^{\alpha}(\gamma^{m}\lambda)_{\alpha}(P_{m}+\partial_{\sigma}X_{m})-\tfrac{1}{2}\hat{L}^{\hat{\alpha}}(\gamma^{m}\hat{\lambda})_{\hat{\alpha}}(P_{m}-\partial_{\sigma}X_{m})\}.\label{eq:Berkovitsaction}
\end{multline}
Here, $\tau$ and $\sigma$ denote the worldsheet coordinates and
hatted and unhatted spinors are related to the usual left and right-moving
variables. The Lagrange multipliers $\{L^{\alpha},\hat{L}^{\hat{\alpha}}\}$
impose the twistor-like constraints and $\{K_{\alpha},\hat{K}_{\hat{\alpha}}\}$,
as a consequence of the gauge algebra, impose the constraints $\nabla_{\sigma}\lambda^{\alpha}=\hat{\nabla}_{\sigma}\hat{\lambda}^{\hat{\alpha}}=0$.
The covariant derivatives $\nabla$ and $\hat{\nabla}$ are composed
by gauge fields, $A$ and $\hat{A}$, associated to local scale symmetries
for $\lambda^{\alpha}$ and $\hat{\lambda}^{\hat{\alpha}}$, which
effectively convert them into projective pure spinors. Although not
explicitly, $S_{B}$ is invariant under worldsheet reparametrization. 

One of the fundamental ideas introduced in these works \cite{Berkovits:2011gh,Berkovits:2014aia,Berkovits:2015yra}
is that worldsheet reparametrization is a redundant gauge symmetry
(in the sense that it can be removed by a gauge-for-gauge transformation).
With this in mind, the action \eqref{eq:Berkovitsaction} is already
incomplete as the absence of a kinetic term for the gauge fields of
the scaling symmetry prevents the existence of a gauge-for-gauge symmetry
connecting the twistor-like constraints and reparametrization. Aside
from this fact, the gauge symmetries due to the pure spinor constraint
imply a constrained ghost system associated to the twistor-like constraint,
such that the gauge fixed action cannot be spacetime supersymmetric.
Observe that the action \eqref{eq:Berkovitsaction} is invariant under
the gauge transformation 
\begin{equation}
\delta L^{\alpha}=f\lambda^{\alpha}+f^{mn}(\gamma_{mn}\lambda)^{\alpha},
\end{equation}
where $\{f,f_{mn}\}$ are the gauge parameters and $\gamma^{mn}=\tfrac{1}{2}(\gamma^{m}\gamma^{n}-\gamma^{n}\gamma^{m})$.
In the gauge fixing procedure, this gauge symmetry appears as a constraint
on the antighost of the twistor-like symmetry, $\pi_{\alpha}$, given
by
\begin{equation}
(\lambda\gamma^{m}\gamma^{n}\pi)=0.\label{eq:BAconstraint-1}
\end{equation}
In turn, it implies that only five components of the associated ghost,
$\theta^{\alpha}$, are physical. Therefore, the phase space of the
action \eqref{eq:Berkovitsaction} should be extended if the twistor-like
constraint is to be part of a spacetime supersymmetric theory.

It is interesting to note, however, that all these features were in
some sense convergent, leading to important concepts that seem to
be connected to the gauge structure behind the pure spinor
superstring, in particular the role of worldsheet reparametrization
and the emergence of spacetime supersymmetry from the ghost sector.

\subsection*{The extended action}

A simple way to understand the physical meaning of the twistor-like
constraint is to look at its worldline version, as the massless particle
can be viewed as the zero length limit of the string. Consider the
constraint equation \eqref{eq:TLconstraint}, but now with a projective
pure spinor. In a Wick-rotated construction, the $SO(10)$ spinor
$H_{\alpha}$ can be decomposed in terms of $U(5)$ components such
that
\begin{equation}
\bar{p}^{a}+\gamma^{ab}p_{b}=0\label{eq:TLU5}
\end{equation}
correspond to the independent components of $H_{\alpha}=0$. Here,
$a=1,\ldots,5$ denotes $U(5)$ vector indices, $P_{m}=\{\bar{p}^{a},p_{a}\}$
and $\gamma^{ab}=-\gamma^{ba}$ corresponds to the $U(5)$ parametrization
of the projective pure spinor. Equation \eqref{eq:TLU5} has a clear
interpretation as any solution of the massless constraint $P^{m}P_{m}=0$
can be put in this form for a dynamical $\gamma^{ab}$. 

The difference between the covariant form \eqref{eq:TLconstraint}
and \eqref{eq:TLU5} is basically the scale symmetry introduced by
Berkovits in \cite{Berkovits:2015yra}. Both of them were recently
investigated by the author in \cite{Jusinskas:2018uci}. In order
to obtain the pure spinor superparticle from first principles, Berkovits'
model was then extended with a constrained anticommuting spinor together
with an additional fermionic gauge symmetry. Now, this idea will be
generalized to the worldsheet with the proposal of the action
\begin{eqnarray}
S_{0} & = & \int d\tau d\sigma\{P_{m}\partial_{\tau}X^{m}-\tfrac{1}{4\mathcal{T}}(L\gamma_{m}\lambda)(P^{m}+\mathcal{T}\partial_{\sigma}X^{m})-\tfrac{1}{4\mathcal{T}}(\hat{L}\gamma_{m}\hat{\lambda})(P^{m}-\mathcal{T}\partial_{\sigma}X^{m})\}\nonumber \\
 &  & +\int d\tau d\sigma\{w_{\alpha}^{i}\nabla_{i}\lambda^{\alpha}+B\epsilon^{ij}\partial_{i}A_{j}+p_{\alpha}^{i}\partial_{i}\xi^{\alpha}+\chi_{i}\lambda^{\alpha}p_{\alpha}^{i}-\Sigma\epsilon^{ij}\nabla_{i}\chi_{j}\}\nonumber \\
 &  & +\int d\tau d\sigma\{\hat{w}_{\hat{\alpha}}^{i}\hat{\nabla}_{i}\hat{\lambda}^{\hat{\alpha}}+\hat{B}\epsilon^{ij}\partial_{i}\hat{A}_{j}+\hat{p}_{\hat{\alpha}}^{i}\partial_{i}\hat{\xi}^{\hat{\alpha}}+\hat{\chi}_{i}\hat{\lambda}^{\hat{\alpha}}\hat{p}_{\hat{\alpha}}^{i}-\hat{\Sigma}\epsilon^{ij}\hat{\nabla}_{i}\hat{\chi}_{j}\}.\label{eq:Jusinskasaction}
\end{eqnarray}
Here, $\{i,j\}$ denote the worldsheet directions $\tau$ and $\sigma$,
and $\mathcal{T}$ is the string tension. The spinors $\xi^{\alpha}$
and $\hat{\xi}^{\hat{\alpha}}$, with conjugates $p_{\alpha}^{i}$
and $\hat{p}_{\hat{\alpha}}^{i}$, satisfy
\begin{equation}
(\lambda\gamma^{m}\xi)=(\hat{\lambda}\gamma^{m}\hat{\xi})=0.
\end{equation}
The motivation for introducing the constrained spinors $\xi^{\alpha}$
and $\hat{\xi}^{\hat{\alpha}}$ is that they should suplement the
degrees of freedom from the ghosts of the twistor-like symmetry, combining
into the superpartners of $X^{m}$. The fermionic symmetry in $S_{0}$
is generated by the current $\lambda^{\alpha}p_{\alpha}^{i}$, with
Lagrange multiplier $\chi_{i}$. The fields $B$ and $\Sigma$ effectively
work as conjugates to the gauge field of the scale symmetry ($A_{i}$)
and its fermionic partner $(\chi_{i}$), respectively. They can be
interpreted as Lagrange multipliers for a zero curvature condition
on the gauge fields, $\epsilon^{ij}\partial_{i}A_{j}=\epsilon^{ij}\nabla_{i}\chi_{j}=0$,
with $\epsilon^{ij}=-\epsilon^{ji}$ ($\epsilon_{ij}=-\epsilon_{ji}$)
and $\epsilon^{\sigma\tau}=\epsilon_{\tau\sigma}=1$. The hatted variables
have an analogous description.

The reparametrization invariant action \eqref{eq:Jusinskasaction}
has a rich gauge structure. Its gauge algebra is onshell reducible
and reparametrization symmetry can be consistently overlooked in the
quantization process due to the simple form of the gauge-for-gauge
symmetries. Still, the Batalin-Vilkovisky formalism seems to be the
most adequate for the quantization of this model, because it provides
a systematic way to analyze the gauge symmetries related to the pure
spinor constraints. The extra fermionic symmetry of the action \eqref{eq:Jusinskasaction}
does not have a clear physical interpretation but it is ultimately
related to the emergence of spacetime supersymmetry in the gauge fixed
action, which can be cast as
\begin{equation}
S=\int d\tau d\sigma\{\tfrac{\mathcal{T}}{2}\partial_{+}X_{m}\partial_{-}X^{m}+w_{\alpha}\partial_{-}\lambda^{\alpha}+p_{\alpha}\partial_{-}\theta^{\alpha}+\hat{w}_{\hat{\alpha}}\partial_{+}\hat{\lambda}^{\hat{\alpha}}+\hat{p}_{\hat{\alpha}}\partial_{+}\hat{\theta}^{\hat{\alpha}}\},\label{eq:Jusinskasfixed}
\end{equation}
where $\partial_{\pm}=\partial_{\tau}\pm\partial_{\sigma}$, corresponding
to the usual superstring action in the pure spinor formalism. In this
action, $\theta^{\alpha}$ and $\hat{\theta}^{\hat{\alpha}}$, the
superpartners of the target space coordinate $X^{m}$, are composed
by the ghosts associated to the twistor-like symmetry complemented
with the constrained spinors $\xi^{\alpha}$ and $\hat{\xi}^{\hat{\alpha}}$.
This composition is only possible due to the existence of the scalar
ghosts $\gamma$ and $\hat{\gamma}$ associated to the new fermionic
symmetry in \eqref{eq:Jusinskasaction}. They are assumed to be everywhere
non-vanishing in the worldsheet, effectively acting as ghost number
twisting operators and turning all the spacetime spinors neutral under
scale transformations. This is the mechanism behind the conversion
of the pure spinors $\lambda^{\alpha}$ and $\hat{\lambda}^{\hat{\alpha}}$
into ghost variables.

\subsection*{Plan of the paper}

Section \ref{sec:PSsuperstring} presents in detail a first principles
derivation of the pure spinor superstring. Subsection \ref{subsec:Polyakov1st}
quickly reviews the first order formulation of the Polyakov action
and subsection \ref{subsec:twistor1st} describes the connection between
the twistor-like constraints and worldsheet reparametrization, motivating
Berkovits' model \cite{Berkovits:2015yra}. In subsection \ref{subsec:A-new-model},
the action \eqref{eq:Jusinskasaction} is introduced with a thorough
analysis of its gauge and gauge-for-gauge symmetries. Subsection \ref{subsec:PSmasteraction}
describes the construction of the pure spinor master action, while
subsection \ref{subsec:Gauge-fixing} presents its gauge fixing. Subsection
\ref{subsec:Fieldred} closes the section with the description of
the emergence of spacetime supersymmetry leading to the pure spinor
superstring. Section \ref{sec:Final-remarks} contains some concluding
remarks and possible directions to follow.

Throughout this work, special attention has been devoted to displaying
auxiliary equations in the intermediate steps, so most computations
can be promptly reproduced. The appendices consist of several complements
and by-products of the main text. Appendix \ref{sec:Partial-gauge-fixing}
shows how the partial gauge fixing of the pure spinor symmetries leads
to the unconstrained spinor $\theta^{\alpha}$, the superpartner of
the target space coordinates $X^{m}$. Appendix \ref{sec:decoupledsector}
presents some properties of the $U(1)_{R}\times U(1)_{L}$ sector
constituted by the ghost fields related to the scaling and fermionic
symmetries, respectively with parameters $\{\Omega,\hat{\Omega}\}$
and $\{\gamma,\hat{\gamma}\}$. Appendix \ref{sec:NMBC} describes
a non-minimal pure spinor superstring with fundamental $(b,c)$ ghosts,
as suggested in \cite{Berkovits:2014aia}. Finally, appendix \ref{sec:sectorized}
presents the sectorized and the ambitwistor string in the pure spinor
formalism as coming from a singular gauge fixing of the action \eqref{eq:Jusinskasaction}.

\section{The pure spinor superstring\label{sec:PSsuperstring}}

In this section, the connection between worldsheet reparametrization
and the twistor-like constraint will be investigated, with a concrete
proposal for the underlying gauge theory of the pure spinor superstring. 

\subsection{Review of the Polyakov action in the first order formalism\label{subsec:Polyakov1st}}

The Polyakov action is given by
\begin{equation}
S_{P}=\frac{\mathcal{T}}{2}\int d^{2}\sigma\sqrt{-g}\{g^{ij}\partial_{i}X^{m}\partial_{j}X_{m}\},
\end{equation}
where $\mathcal{T}$ is the string tension, $g^{ij}$ is the worldsheet
metric (with inverse $g_{ij}$) and $g=\det(g_{ij})$. When expanded
in components ($i=\tau,\sigma$), with $\tau$ denoting the worldsheet
time and $\sigma$ parametrizing the string length, $S_{P}$ is rewritten
as
\begin{equation}
S_{P}=\frac{\mathcal{T}}{2}\int d\tau d\sigma\,\sqrt{-g}\{g^{\tau\tau}\partial_{\tau}X^{m}\partial_{\tau}X_{m}+2g^{\tau\sigma}\partial_{\tau}X^{m}\partial_{\sigma}X_{m}+g^{\sigma\sigma}\partial_{\sigma}X^{m}\partial_{\sigma}X_{m}\}.
\end{equation}
The canonical conjugate of the target-space coordinate $X^{m}$ is
easily determined to be
\begin{eqnarray}
P_{m} & \equiv & \frac{\delta S}{\delta\partial_{\tau}X^{m}},\nonumber \\
 & = & \mathcal{T}\sqrt{-g}\{g^{\tau\tau}\partial_{\tau}X_{m}+g^{\tau\sigma}\partial_{\sigma}X_{m}\},\label{eq:Pmeom}
\end{eqnarray}
leading to the Hamiltonian
\begin{eqnarray}
H & \equiv & P_{m}\partial_{\tau}X^{m}-\tfrac{\mathcal{T}}{2}\sqrt{-g}g^{ij}\partial_{i}X^{m}\partial_{j}X_{m},\nonumber \\
 & = & \frac{1}{2\mathcal{T}g^{\tau\tau}\sqrt{-g}}(P_{m}P^{m}+\mathcal{T}^{2}\partial_{\sigma}X_{m}\partial_{\sigma}X^{m})-\frac{g^{\tau\sigma}}{g^{\tau\tau}}P_{m}\partial_{\sigma}X^{m}.
\end{eqnarray}

In the first order formulation, the Polyakov action takes the form
\begin{equation}
\tilde{S}_{P}=\int d\tau d\sigma\{P_{m}\partial_{\tau}X^{m}-\frac{1}{2\mathcal{T}g^{\tau\tau}\sqrt{-g}}(P_{m}P^{m}+\mathcal{T}^{2}\partial_{\sigma}X_{m}\partial_{\sigma}X^{m})+\frac{g^{\tau\sigma}}{g^{\tau\tau}}P_{m}\partial_{\sigma}X^{m}\},
\end{equation}
which is equivalent onshell to $S_{P}$, \emph{cf}. equation \eqref{eq:Pmeom}.
Observe that the only dependence on the worldsheet metric appears
now in the form of Lagrange multipliers. In fact, by defining the
Weyl invariant operators
\begin{equation}
e_{\pm}\equiv\frac{1}{g^{\tau\tau}\sqrt{-g}}\mp\frac{g^{\tau\sigma}}{g^{\tau\tau}},\label{eq:e+-tometric}
\end{equation}
the action $\tilde{S}_{P}$ is more symmetrically rewritten as
\begin{multline}
\tilde{S}_{P}=\int d\tau d\sigma\big\{ P_{m}\partial_{\tau}X^{m}-\tfrac{1}{4\mathcal{T}}e_{+}(P_{m}+\mathcal{T}\partial_{\sigma}X_{m})(P^{m}+\mathcal{T}\partial_{\sigma}X^{m})\\
-\tfrac{1}{4\mathcal{T}}e_{-}(P_{m}-\mathcal{T}\partial_{\sigma}X_{m})(P^{m}-\mathcal{T}\partial_{\sigma}X^{m})\big\}.\label{eq:Pol1st}
\end{multline}

The equations of motion for $P_{m}$, $X^{m}$ and $e_{\pm}$ are
respectively given by\begin{subequations}
\begin{eqnarray}
\partial_{\tau}X^{m}-\tfrac{1}{2\mathcal{T}}[e_{+}(P^{m}+\mathcal{T}\partial_{\sigma}X^{m})+e_{-}(P^{m}-\mathcal{T}\partial_{\sigma}X^{m})] & = & 0\\
\partial_{\tau}P_{m}-\tfrac{1}{2}\partial_{\sigma}[e_{+}(P_{m}+\mathcal{T}\partial_{\sigma}X_{m})-e_{-}(P_{m}-\mathcal{T}\partial_{\sigma}X_{m})] & = & 0,\\
(P_{m}\pm\mathcal{T}\partial_{\sigma}X_{m})(P^{m}\pm\mathcal{T}\partial_{\sigma}X^{m}) & = & 0,
\end{eqnarray}
\end{subequations} and the action \eqref{eq:Pol1st} is invariant
under the gauge transformations\begin{subequations}\label{eq:Pol1stgauge}
\begin{eqnarray}
\delta X^{m} & = & c^{i}\partial_{i}X^{m}-\tfrac{1}{2\mathcal{T}}a^{+}(P^{m}+\mathcal{T}\partial_{\sigma}X^{m})-\tfrac{1}{2\mathcal{T}}a^{-}(P^{m}-\mathcal{T}\partial_{\sigma}X^{m}),\\
\delta P_{m} & = & \partial_{i}(c^{i}P_{m})-P_{m}\partial_{\tau}c^{\tau}+\tfrac{1}{2}\partial_{\sigma}c^{\tau}[(P_{m}+\mathcal{T}\partial_{\sigma}X_{m})e_{+}-(P_{m}-\mathcal{T}\partial_{\sigma}X_{m})e_{-}]\nonumber \\
 &  & -\tfrac{1}{2}\partial_{\sigma}[a^{+}(P^{m}+\mathcal{T}\partial_{\sigma}X^{m})-a^{-}(P^{m}-\mathcal{T}\partial_{\sigma}X^{m})],\\
\delta e_{+} & = & \partial_{i}(c^{i}e_{+})-2(\partial_{\sigma}c^{\sigma})e_{+}+\partial_{\tau}c^{\sigma}-e_{+}^{2}\partial_{\sigma}c^{\tau}-\partial_{\tau}a^{+}-a^{+}\partial_{\sigma}e_{+}+e_{+}\partial_{\sigma}a^{+},\\
\delta e_{-} & = & \partial_{i}(c^{i}e_{-})-2(\partial_{\sigma}c^{\sigma})e_{-}-\partial_{\tau}c^{\sigma}+e_{-}^{2}\partial_{\sigma}c^{\tau}-\partial_{\tau}a^{-}+a^{-}\partial_{\sigma}e_{-}-e_{-}\partial_{\sigma}a^{-}.
\end{eqnarray}
\end{subequations}In addition to worldsheet reparametrization symmetry,
with parameter $c^{i}$, the action $\tilde{S}_{P}$ is also invariant
under the gauge transformations parametrized by $a^{\pm}$. However,
these gauge symmetries are not irreducible. To see this, consider
the gauge-for-gauge transformations\begin{subequations}
\begin{eqnarray}
\delta^{'}c^{i} & = & \phi^{i},\\
\delta^{'}a^{\pm} & = & \phi^{\tau}e_{\pm}\pm\phi^{\sigma},
\end{eqnarray}
\end{subequations}with parameter $\phi^{i}$. It is then straightforward
to show that the gauge transformations \eqref{eq:Pol1stgauge} are
invariant up to equations of motion: \begin{subequations}
\begin{eqnarray}
\delta^{'}[\delta X^{m}] & = & \phi^{\tau}\big\{\partial_{\tau}X^{m}-\tfrac{1}{2\mathcal{T}}[e_{+}(P^{m}+\mathcal{T}\partial_{\sigma}X^{m})+e_{-}(P^{m}-\mathcal{T}\partial_{\sigma}X^{m})]\big\},\\
\delta^{'}[\delta P_{m}] & = & \phi^{\tau}\big\{\partial_{\tau}P_{m}-\tfrac{1}{2}\partial_{\sigma}[e_{+}(P_{m}+\mathcal{T}\partial_{\sigma}X_{m})-e_{-}(P_{m}-\mathcal{T}\partial_{\sigma}X_{m})]\big\},\\
\delta^{'}[\delta e_{+}] & = & 0,\\
\delta^{'}[\delta e_{-}] & = & 0.
\end{eqnarray}
\end{subequations}Therefore, worldsheet reparametrization is equivalent
to the symmetries generated by
\begin{eqnarray}
H^{\pm} & \equiv & (P_{m}\pm\mathcal{T}\partial_{\sigma}X_{m})(P^{m}\pm\mathcal{T}\partial_{\sigma}X^{m}).\label{eq:H+-}
\end{eqnarray}

\subsection{The twistor-like constraint in the first order formalism\label{subsec:twistor1st}}

In \cite{Berkovits:2015yra}, Berkovits proposed the twistor-like
constraints\begin{subequations}\label{eq:twistorconstraints}
\begin{eqnarray}
H_{\alpha} & \equiv & (P_{m}+\mathcal{T}\partial_{\sigma}X_{m})(\gamma^{m}\lambda)_{\alpha},\\
\hat{H}_{\hat{\alpha}} & \equiv & (P_{m}-\mathcal{T}\partial_{\sigma}X_{m})(\gamma^{m}\hat{\lambda})_{\hat{\alpha}},
\end{eqnarray}
\end{subequations}as part of the fundamental gauge algebra behind
the pure spinor superstring, where $\lambda^{\alpha}$ and $\hat{\lambda}^{\hat{\alpha}}$
are bosonic spinors satisfying the pure spinor condition\begin{subequations}\label{eq:PSconstraint}
\begin{eqnarray}
(\lambda\gamma^{m}\lambda) & = & 0,\\
(\hat{\lambda}\gamma^{m}\hat{\lambda}) & = & 0.
\end{eqnarray}
\end{subequations}The key idea here is that $H^{\pm}$ in \eqref{eq:H+-}
can be rewritten as\begin{subequations}\label{eq:H+-twistor}
\begin{eqnarray}
H^{+} & = & (P_{m}+\mathcal{T}\partial_{\sigma}X_{m})\frac{(\Lambda\gamma^{m})^{\alpha}}{(\Lambda\lambda)}H_{\alpha},\\
H^{-} & = & (P_{m}-\mathcal{T}\partial_{\sigma}X_{m})\frac{(\gamma^{m}\hat{\Lambda})^{\hat{\alpha}}}{(\hat{\Lambda}\hat{\lambda})}\hat{H}_{\hat{\alpha}},
\end{eqnarray}
\end{subequations}for any constant $\Lambda_{\alpha}$ and $\hat{\Lambda}_{\hat{\alpha}}$
with non-vanishing $(\Lambda\lambda)$ and $(\hat{\Lambda}\hat{\lambda})$. 

The first order Polyakov's action \eqref{eq:Pol1st} can be covariantly
modified with the introduction of the twistor-like constraints \eqref{eq:twistorconstraints}.
In order to do that, $\lambda^{\alpha}$ and $\hat{\lambda}^{\hat{\alpha}}$
have to be made dynamical. In addition, Berkovits proposed in \cite{Berkovits:2015yra}
the use of projective pure spinors which can be achieved by endowing
$\lambda^{\alpha}$ and $\hat{\lambda}^{\hat{\alpha}}$ with a scaling
symmetry. The resulting action is
\begin{multline}
S_{B}^{'}=\int d\tau d\sigma\{P_{m}\partial_{\tau}X^{m}+w_{\alpha}^{i}\partial_{i}\lambda^{\alpha}+\hat{w}_{\hat{\alpha}}^{i}\partial_{i}\hat{\lambda}^{\hat{\alpha}}+A_{i}\lambda^{\alpha}w_{\alpha}^{i}+\hat{A}_{i}\hat{\lambda}^{\hat{\alpha}}\hat{w}_{\hat{\alpha}}^{i}\}\\
-\tfrac{1}{4\mathcal{T}}\int d\tau d\sigma\Big\{(L\gamma_{m}\lambda)(P^{m}+\mathcal{T}\partial_{\sigma}X^{m})+e_{+}(P_{m}+\mathcal{T}\partial_{\sigma}X_{m})(P^{m}+\mathcal{T}\partial_{\sigma}X^{m})\\
+(\hat{L}\gamma_{m}\hat{\lambda})(P^{m}-\mathcal{T}\partial_{\sigma}X^{m})+e_{-}(P_{m}-\mathcal{T}\partial_{\sigma}X_{m})(P^{m}-\mathcal{T}\partial_{\sigma}X^{m})\Big\},
\end{multline}
where $\{L^{\alpha},\hat{L}^{\hat{\alpha}}\}$ are the Lagrange multipliers
for the constraints \eqref{eq:twistorconstraints} and $\{A_{i},\hat{A}_{i}\}$
are the gauge fields for the scaling symmetry generated by $\lambda^{\alpha}w_{\alpha}^{i}$
and $\hat{\lambda}^{\hat{\alpha}}\hat{w}_{\hat{\alpha}}^{i}$. Due
to the identification \eqref{eq:H+-twistor}, a field shift in $L^{\alpha}$
and $\hat{L}^{\hat{\alpha}}$ can absorb $e_{+}$ and $e_{-}$, leading
to Berkovits' action,
\begin{eqnarray}
S_{B} & = & \int d\tau d\sigma\{P_{m}\partial_{\tau}X^{m}+w_{\alpha}^{i}\nabla_{i}\lambda^{\alpha}+\hat{w}_{\hat{\alpha}}^{i}\hat{\nabla}_{i}\hat{\lambda}^{\hat{\alpha}}\}\nonumber \\
 &  & -\tfrac{1}{4\mathcal{T}}\int d\tau d\sigma\{(L\gamma_{m}\lambda)(P^{m}+\mathcal{T}\partial_{\sigma}X^{m})+(\hat{L}\gamma_{m}\hat{\lambda})(P^{m}-\mathcal{T}\partial_{\sigma}X^{m})\}.\label{eq:Berkovitsactioncovariant}
\end{eqnarray}
The gauge fields $\{A_{i},\hat{A}_{i}\}$ now appear through the covariant
derivatives $\{\nabla_{i},\hat{\nabla}_{i}\}$. Observe also that
all the dependence on the worldsheet metric is concentrated in the
Lagrange multipliers $\{L^{\alpha},\hat{L}^{\hat{\alpha}}\}$ and
this has to be taken into account during the gauge fixing process.

\subsection{A new model with constrained anticommuting spinors\label{subsec:A-new-model}}

Based on the worldline results of \cite{Jusinskas:2018uci}, it is
straightforward to generalize the action \eqref{eq:Berkovitsactioncovariant}
to
\begin{eqnarray}
S_{0} & = & \int d\tau d\sigma\{P_{m}\partial_{\tau}X^{m}+w_{\alpha}^{i}\nabla_{i}\lambda^{\alpha}+\hat{w}_{\hat{\alpha}}^{i}\hat{\nabla}_{i}\hat{\lambda}^{\hat{\alpha}}\}\nonumber \\
 &  & -\tfrac{1}{4\mathcal{T}}\int d\tau d\sigma\{(L\gamma_{m}\lambda)(P^{m}+\mathcal{T}\partial_{\sigma}X^{m})+(\hat{L}\gamma_{m}\hat{\lambda})(P^{m}-\mathcal{T}\partial_{\sigma}X^{m})\}\nonumber \\
 &  & +\int d\tau d\sigma\{p_{\alpha}^{i}\partial_{i}\xi^{\alpha}+B\epsilon^{ij}\partial_{i}A_{j}+\hat{p}_{\hat{\alpha}}^{i}\partial_{i}\hat{\xi}^{\hat{\alpha}}+\hat{B}\epsilon^{ij}\partial_{i}\hat{A}_{j}\}\nonumber \\
 &  & +\int d\tau d\sigma\{\chi_{i}\lambda^{\alpha}p_{\alpha}^{i}-\Sigma\epsilon^{ij}\nabla_{i}\chi_{j}+\hat{\chi}_{i}\hat{\lambda}^{\hat{\alpha}}\hat{p}_{\hat{\alpha}}^{i}-\hat{\Sigma}\epsilon^{ij}\hat{\nabla}_{i}\hat{\chi}_{j}\}.\label{eq:newaction}
\end{eqnarray}
There are two guiding principles that led to the proposed action $S_{0}$,
(1) the extension of the phase space with the inclusion of constrained
anticommuting spinors, $\xi^{\alpha}$ and $\hat{\xi}^{\hat{\alpha}}$,
satisfying\begin{subequations}\label{eq:SPSconstraint}
\begin{eqnarray}
(\lambda\gamma^{m}\xi) & = & 0,\\
(\hat{\lambda}\gamma^{m}\hat{\xi}) & = & 0,
\end{eqnarray}
\end{subequations}together with two fermionic symmetries generated
by $\lambda^{\alpha}p_{\alpha}^{i}$ and $\hat{\lambda}^{\hat{\alpha}}\hat{p}_{\hat{\alpha}}^{i}$;
and (2) the introduction of zero curvature conditions on the gauge
fields $\{A_{i},\chi_{i},\hat{A}_{i},\hat{\chi}_{i}\}$ through the
Lagrange multipliers $\{B,\Sigma,\hat{B},\hat{\Sigma}\}$, which enables
the extension of the gauge algebra of the model with the inclusion
of gauge-for-gauge symmetries connecting worldsheet reparametrization
and the twistor-like symmetries.

The equations of motion obtained from the action \eqref{eq:newaction}
can be summarized as\begin{subequations}\label{eq:eomPS}
\begin{eqnarray}
\partial_{\tau}X^{m}-\tfrac{1}{4\mathcal{T}}(L\gamma^{m}\lambda)-\tfrac{1}{4\mathcal{T}}(\hat{L}\gamma^{m}\hat{\lambda}) & = & 0,\\
\partial_{\tau}P_{m}-\tfrac{1}{4}\partial_{\sigma}(L\gamma_{m}\lambda)+\tfrac{1}{4}\partial_{\sigma}(\hat{L}\gamma_{m}\hat{\lambda}) & = & 0,\\
\nabla_{i}\lambda^{\alpha} & = & 0,\\
\nabla_{i}w_{\alpha}^{i}+\tfrac{1}{4\mathcal{T}}(P^{m}+\mathcal{T}\partial_{\sigma}X^{m})(\gamma_{m}L)_{\alpha}-\chi_{i}p_{\alpha}^{i} & = & 0,\\
\lambda^{\alpha}w_{\alpha}^{i}+\epsilon^{ij}(\Sigma\chi_{j}+\partial_{j}B) & = & 0,\\
\partial_{i}\xi^{\alpha}-\chi_{i}\lambda^{\alpha} & = & 0,\\
\partial_{i}p_{\alpha}^{i} & = & 0,\\
\epsilon^{ij}\partial_{i}A_{j} & = & 0,\\
(P^{m}+\mathcal{T}\partial_{\sigma}X^{m})(\gamma_{m}\lambda)_{\alpha} & = & 0,\\
\lambda^{\alpha}p_{\alpha}^{i}+\epsilon^{ij}\nabla_{j}\Sigma & = & 0,\\
\epsilon^{ij}\nabla_{i}\chi_{j} & = & 0,\\
\hat{\nabla}_{i}\hat{\lambda}^{\hat{\alpha}} & = & 0,\\
\hat{\nabla}_{i}\hat{w}_{\hat{\alpha}}^{i}+\tfrac{1}{4\mathcal{T}}(P^{m}-\mathcal{T}\partial_{\sigma}X^{m})(\gamma_{m}\hat{L})_{\hat{\alpha}}-\hat{\chi}_{i}\hat{p}_{\hat{\alpha}}^{i} & = & 0,\\
\hat{\lambda}^{\hat{\alpha}}\hat{w}_{\hat{\alpha}}^{i}+\epsilon^{ij}(\hat{\Sigma}\hat{\chi}_{j}+\partial_{j}\hat{B}) & = & 0,\\
\partial_{i}\hat{\xi}^{\hat{\alpha}}-\hat{\chi}_{i}\hat{\lambda}^{\hat{\alpha}} & = & 0,\\
\partial_{i}\hat{p}_{\hat{\alpha}}^{i} & = & 0,\\
\epsilon^{ij}\partial_{i}\hat{A}_{j} & = & 0,\\
(P^{m}-\mathcal{T}\partial_{\sigma}X^{m})(\gamma_{m}\hat{\lambda})_{\alpha} & = & 0,\\
\hat{\lambda}^{\hat{\alpha}}\hat{p}_{\hat{\alpha}}^{i}+\epsilon^{ij}\hat{\nabla}_{j}\hat{\Sigma} & = & 0,\\
\epsilon^{ij}\hat{\nabla}_{i}\hat{\chi}_{j} & = & 0.
\end{eqnarray}
\end{subequations}

Due to the constraints \eqref{eq:PSconstraint} and \eqref{eq:SPSconstraint},
the action $S_{0}$ is invariant under
\begin{equation}
\begin{array}{rclcrcl}
\delta w_{\alpha}^{i} & = & d_{m}^{i}(\gamma^{m}\lambda)_{\alpha}+e_{m}^{i}(\gamma^{m}\xi)_{\alpha}, &  & \delta\hat{w}_{\hat{\alpha}}^{i} & = & \hat{d}_{m}^{i}(\gamma^{m}\hat{\lambda})_{\hat{\alpha}}+\hat{e}_{m}^{i}(\gamma^{m}\hat{\xi})_{\hat{\alpha}},\\
\delta p_{\alpha}^{i} & = & e_{m}^{i}(\gamma^{m}\lambda)_{\alpha}, &  & \delta\hat{p}_{\hat{\alpha}}^{i} & = & \hat{e}_{m}^{i}(\gamma^{m}\hat{\lambda})_{\hat{\alpha}},\\
\delta L^{\alpha} & = & f\lambda^{\alpha}+f_{mn}(\gamma^{mn}\lambda)^{\alpha}+g\xi^{\alpha}, &  & \delta\hat{L}^{\hat{\alpha}} & = & \hat{f}\hat{\lambda}^{\hat{\alpha}}+\hat{f}_{mn}(\gamma^{mn}\hat{\lambda})^{\hat{\alpha}}+\hat{g}\xi^{\hat{\alpha}},
\end{array}\label{eq:prePSsymetries}
\end{equation}
where $d_{m}$, $e_{m}$, $f$, $f_{mn}$, $g$ (hatted and unhatted)
are local parameters. These gauge transformations have a special role
in the formalism and will be called pure spinor symmetries. The other
gauge symmetries of the model can be summarized by:
\begin{enumerate}
\item Worldsheet reparametrization, with parameter $c^{i}=\{c^{\tau},c^{\sigma}\}$.
Although the transformations of $P_{m}$, $L^{\alpha}$ and $\hat{L}^{\hat{\alpha}}$
are nontrivial,
\begin{equation}
\begin{array}{rcl}
\delta P_{m} & = & \partial_{\sigma}(c^{\sigma}P_{m})+c^{\tau}\partial_{\tau}P_{m}+\tfrac{1}{4}(L_{+}\gamma_{m}\lambda)(\partial_{\sigma}c^{\tau})-\tfrac{1}{4}(\hat{L}_{-}\gamma_{m}\hat{\lambda})(\partial_{\sigma}c^{\tau}),\\
\delta L^{\alpha} & = & c^{\sigma}\partial_{\sigma}L^{\alpha}+\partial_{\tau}(c^{\tau}L^{\alpha})+\partial_{\tau}c^{\sigma}(P^{m}+\mathcal{T}\partial_{\sigma}X^{m})\frac{(\gamma_{m}\Lambda)^{\alpha}}{(\Lambda\lambda)},\\
\delta\hat{L}^{\hat{\alpha}} & = & c^{\sigma}\partial_{\sigma}\hat{L}^{\hat{\alpha}}+\partial_{\tau}(c^{\tau}\hat{L}^{\hat{\alpha}})-\partial_{\tau}c^{\sigma}(P^{m}-\mathcal{T}\partial_{\sigma}X^{m})\frac{(\gamma_{m}\hat{\Lambda})^{\hat{\alpha}}}{(\hat{\Lambda}\hat{\lambda})},
\end{array}
\end{equation}
all the other fields transform covariantly either as worldsheet scalars
(\emph{e.g.} $\delta X^{m}=c^{i}\partial_{i}X^{m}$), vector densities
(\emph{e.g.} $\delta w_{\alpha}^{i}=\partial_{j}(c^{j}w_{\alpha}^{i})-w_{\alpha}^{j}\partial_{j}c^{i},$)
or 1-forms (\emph{e.g.} $\delta A_{i}=c^{j}\partial_{j}A_{i}+A_{j}\partial_{i}c^{j}$).
\item Particle-like Hamiltonian symmetry, with parameter $a^{\pm}$. This
symmetry is analogous to \eqref{eq:Pol1stgauge} and the transformations
can be cast as
\begin{equation}
\begin{array}{rcl}
\delta X^{m} & = & a^{+}(P^{m}+\mathcal{T}\partial_{\sigma}X^{m})+a^{-}(P^{m}-\mathcal{T}\partial_{\sigma}X^{m}),\\
\delta P_{m} & = & \mathcal{T}\partial_{\sigma}[a^{+}(P^{m}+\mathcal{T}\partial_{\sigma}X^{m})]-\mathcal{T}\partial_{\sigma}[a^{-}(P^{m}-\mathcal{T}\partial_{\sigma}X^{m})],\\
\delta L^{\alpha} & = & 2\mathcal{T}a^{+}\nabla_{\sigma}L^{\alpha}+2\mathcal{T}\partial_{\tau}a^{+}(P^{m}+\mathcal{T}\partial_{\sigma}X^{m})\frac{(\gamma_{m}\Lambda)^{\alpha}}{(\Lambda\lambda)},\\
\delta w_{\alpha}^{\sigma} & = & -\tfrac{1}{2}a^{+}(P^{m}+\mathcal{T}\partial_{\sigma}X^{m})(\gamma_{m}L)_{\alpha},\\
\delta\hat{L}^{\hat{\alpha}} & = & -2\mathcal{T}a^{-}\hat{\nabla}_{\sigma}\hat{L}^{\hat{\alpha}}+2\mathcal{T}\partial_{\tau}a^{-}(P_{m}-\mathcal{T}\partial_{\sigma}X_{m})\frac{(\gamma^{m}\hat{\Lambda})^{\hat{\alpha}}}{(\hat{\lambda}\hat{\Lambda})},\\
\delta\hat{w}_{\hat{\alpha}}^{\sigma} & = & \tfrac{1}{2}a^{-}(P^{m}-\mathcal{T}\partial_{\sigma}X^{m})(\gamma_{m}\hat{L})_{\hat{\alpha}}.
\end{array}
\end{equation}
\item Scaling symmetry, with parameters $\{\Omega,\hat{\Omega}\}$. The
transformations are
\begin{equation}
\begin{array}{rclcrcl}
\delta\lambda^{\alpha} & = & \Omega\lambda^{\alpha}, &  & \delta\hat{\lambda}^{\hat{\alpha}} & = & \hat{\Omega}\hat{\lambda}^{\hat{\alpha}},\\
\delta w_{\alpha}^{i} & = & -\Omega w_{\alpha}^{i}, &  & \delta\hat{w}_{\hat{\alpha}}^{i} & = & -\Omega\hat{w}_{\hat{\alpha}}^{i},\\
\delta A_{i} & = & -\partial_{i}\Omega, &  & \delta\hat{A}_{i} & = & -\partial_{i}\hat{\Omega},\\
\delta L^{\alpha} & = & -\Omega L^{\alpha}, &  & \delta\hat{L}^{\hat{\alpha}} & = & -\hat{\Omega}\hat{L}^{\hat{\alpha}},\\
\delta\chi_{i} & = & -\Omega\chi_{i}, &  & \delta\hat{\chi}_{i} & = & -\hat{\Omega}\hat{\chi}_{i},\\
\delta\Sigma & = & \Omega\Sigma, &  & \delta\hat{\Sigma} & = & \hat{\Omega}\hat{\Sigma}.
\end{array}\label{eq:scalingSYM}
\end{equation}
\item Fermionic gauge symmetry, with gauge fields $\chi_{i}$ and $\hat{\chi}_{i}$
and parameters $\gamma$ and $\hat{\gamma}$, respectively. The action
$S_{0}$ is invariant under the transformations
\begin{equation}
\begin{array}{rclcrcl}
\delta w_{\alpha}^{i} & = & \gamma p_{\alpha}^{i}, &  & \delta\hat{w}_{\hat{\alpha}}^{i} & = & \hat{\gamma}\hat{p}_{\hat{\alpha}}^{i},\\
\delta B & = & \gamma\Sigma, &  & \delta\hat{B} & = & \hat{\gamma}\hat{\Sigma},\\
\delta\xi^{\alpha} & = & \gamma\lambda^{\alpha}, &  & \delta\hat{\xi}^{\hat{\alpha}} & = & \hat{\gamma}\hat{\lambda}^{\hat{\alpha}},\\
\delta\chi_{i} & = & \nabla_{i}\gamma, &  & \delta\chi_{i} & = & \hat{\nabla}_{i}\hat{\gamma}.
\end{array}\label{eq:fermionSYM}
\end{equation}
\item Curl symmetry, with parameters $\{s_{\alpha},\epsilon_{\alpha},\hat{s}_{\hat{\alpha}},\hat{\epsilon}_{\hat{\alpha}}\}$.
By construction, the reparametrization invariant form of the kinetic
terms of the spinors in \eqref{eq:newaction} imply the existence
of gauge transformations given by
\begin{equation}
\begin{array}{rclcrcl}
\delta w_{\alpha}^{i} & = & \epsilon^{ij}(\nabla_{j}s_{\alpha}+\epsilon_{\alpha}\chi_{j}), &  & \delta\hat{w}_{\hat{\alpha}}^{i} & = & \epsilon^{ij}(\hat{\nabla}_{j}\hat{s}_{\hat{\alpha}}+\hat{\epsilon}_{\hat{\alpha}}\hat{\chi}_{j}),\\
\delta B & = & -\lambda^{\alpha}s_{\alpha}, &  & \delta\hat{B} & = & -\hat{\lambda}^{\hat{\alpha}}\hat{s}_{\hat{\alpha}},\\
\delta p_{\alpha}^{i} & = & \epsilon^{ij}\partial_{j}\epsilon_{\alpha}, &  & \delta\hat{p}_{\hat{\alpha}} & = & \epsilon^{ij}\partial_{j}\hat{\epsilon}_{\hat{\alpha}},\\
\delta\Sigma & = & -\lambda^{\alpha}\epsilon_{\alpha}, &  & \delta\hat{\Sigma} & = & -\hat{\lambda}^{\hat{\alpha}}\hat{\epsilon}_{\hat{\alpha}}.
\end{array}\label{eq:curlSYM}
\end{equation}
\item Twistor-like symmetry, with parameters $\theta^{\alpha}$ and $\hat{\theta}^{\hat{\alpha}}$
and gauge transformations
\begin{equation}
\begin{array}{rclcrcl}
\delta X^{m} & = & \tfrac{1}{4T}\big[(\lambda\gamma^{m}\theta)+(\hat{\lambda}\gamma^{m}\hat{\theta})\big], &  & \delta w_{\alpha}^{i} & = & -\delta_{\tau}^{i}\tfrac{1}{4\mathcal{T}}(P_{m}+\mathcal{T}\partial_{\sigma}X^{m})(\theta\gamma^{m})_{\alpha}\\
\delta P_{m} & = & \tfrac{1}{4}\nabla_{\sigma}\big[(\lambda\gamma^{m}\theta)-(\hat{\lambda}\gamma^{m}\hat{\theta})\big], &  &  &  & +\delta_{\sigma}^{i}\tfrac{1}{8\mathcal{T}}(L\gamma_{m}\lambda)(\theta\gamma^{m})_{\alpha},\\
\delta L^{\alpha} & = & \nabla_{\tau}\theta^{\alpha}, &  & \delta\hat{w}_{\hat{\alpha}}^{i} & = & -\delta_{\tau}^{i}\tfrac{1}{4\mathcal{T}}(P_{m}-\mathcal{T}\partial_{\sigma}X^{m})(\hat{\theta}\gamma^{m})_{\hat{\alpha}}\\
\delta\hat{L}^{\hat{\alpha}} & = & \hat{\nabla}_{\tau}\hat{\theta}^{\hat{\alpha}}, &  &  &  & -\delta_{\sigma}^{i}\tfrac{1}{8\mathcal{T}}(\hat{L}\gamma_{m}\hat{\lambda})(\hat{\theta}\gamma^{m})_{\hat{\alpha}}.
\end{array}\label{eq:twistorSYM}
\end{equation}
\end{enumerate}
\

To complete the analysis of the gauge structure of $S_{0}$, consider
the gauge-for-gauge transformations with parameters $\phi^{i}$ and
$\varphi^{\pm}$:
\begin{equation}
\begin{array}{rclcrcl}
\delta^{'}c^{i} & = & \phi^{i}, &  & \delta^{'}\Omega & = & \phi^{i}A_{i},\\
\delta^{'}a^{\pm} & = & \varphi^{\pm}\mp\phi^{\sigma}, &  & \delta^{'}\hat{\Omega} & = & \phi^{i}\hat{A}_{i},\\
\delta^{'}s_{\alpha} & = & \epsilon_{ij}\phi^{j}w_{\alpha}^{i}, &  & \delta^{'}\epsilon_{\alpha} & = & \epsilon_{ij}\phi^{j}p_{\alpha}^{i},\\
\delta^{'}\hat{s}_{\hat{\alpha}} & = & \epsilon_{ij}\phi^{j}\hat{w}_{\hat{\alpha}}^{i}, &  & \delta^{'}\hat{\epsilon}_{\hat{\alpha}} & = & \epsilon_{ij}\phi^{j}\hat{p}_{\hat{\alpha}}^{i},\\
\delta^{'}\gamma & = & -\phi^{i}\chi_{i}, &  & \delta^{'}\theta^{\alpha} & = & -\phi^{\tau}L^{\alpha}-\varphi^{+}(P_{m}+\mathcal{T}\partial_{\sigma}X_{m})\frac{(\gamma^{m}\Lambda)^{\alpha}}{(\Lambda\lambda)},\\
\delta^{'}\hat{\gamma} & = & -\phi^{i}\hat{\chi}_{i}, &  & \delta^{'}\hat{\theta}^{\hat{\alpha}} & = & -\phi^{\tau}\hat{L}^{\hat{\alpha}}-\varphi^{-}(P_{m}-\mathcal{T}\partial_{\sigma}X_{m})\frac{(\gamma^{m}\hat{\Lambda})^{\hat{\alpha}}}{(\hat{\lambda}\hat{\Lambda})}.
\end{array}\label{eq:g4g}
\end{equation}
Up to equations of motion and pure spinor symmetries,\emph{ cf}. equations
\eqref{eq:eomPS} and \eqref{eq:prePSsymetries}, the gauge transformations
listed above are left invariant by \eqref{eq:g4g}:\begin{subequations}
\begin{eqnarray}
\delta^{'}[\delta X^{m}] & = & \phi^{\tau}\Big\{\partial_{\tau}X^{m}-\tfrac{1}{4\mathcal{T}}(\lambda\gamma^{m}L)-\tfrac{1}{4\mathcal{T}}(\hat{\lambda}\gamma^{m}\hat{L})\Big\}\nonumber \\
 &  & +\tfrac{1}{4\mathcal{T}}\varphi^{+}\big\{(P_{n}+\mathcal{T}\partial_{\sigma}X_{n})(\gamma^{n}\lambda)_{\alpha}\big\}\tfrac{(\gamma^{m}\Lambda)^{\alpha}}{(\Lambda\lambda)}\nonumber \\
 &  & +\tfrac{1}{4\mathcal{T}}\varphi^{-}\big\{(P_{n}-\mathcal{T}\partial_{\sigma}X_{n})(\gamma^{n}\hat{\lambda})_{\hat{\alpha}}\big\}\tfrac{(\gamma^{m}\hat{\Lambda})^{\alpha}}{(\hat{\Lambda}\hat{\lambda})},\\
\delta^{'}[\delta P_{m}] & = & \phi^{\tau}\Big\{\partial_{\tau}P_{m}-\tfrac{1}{4}\partial_{\sigma}(\lambda\gamma_{m}L)+\tfrac{1}{4}\partial_{\sigma}(\hat{\lambda}\gamma_{m}\hat{L})\Big\}\nonumber \\
 &  & +\tfrac{1}{4}\nabla_{\sigma}\Big[\varphi^{+}\big\{(P_{n}+\mathcal{T}\partial_{\sigma}X_{n})(\gamma^{n}\lambda)_{\alpha}\big\}\tfrac{(\gamma^{m}\Lambda)^{\alpha}}{(\Lambda\lambda)}\Big]\nonumber \\
 &  & -\tfrac{1}{4}\hat{\nabla}_{\sigma}\Big[\varphi^{-}\big\{(P_{n}-\mathcal{T}\partial_{\sigma}X_{n})(\gamma^{n}\hat{\lambda})_{\hat{\alpha}}\big\}\tfrac{(\gamma^{m}\hat{\Lambda})^{\hat{\alpha}}}{(\hat{\Lambda}\hat{\lambda})}\Big],\\
\delta^{'}[\delta\lambda^{\alpha}] & = & \phi^{i}\{\nabla_{i}\lambda^{\alpha}\},\\
\delta^{'}[\delta w_{\alpha}^{i}] & = & \phi^{i}\big\{\nabla_{j}w_{\alpha}^{j}+\tfrac{1}{4T}(P_{m}+\mathcal{T}\partial_{\sigma}X_{m})(\gamma^{m}L)_{\alpha}-\chi_{j}p_{\alpha}^{j}\big\}\nonumber \\
 &  & +\tfrac{1}{16\mathcal{T}}\delta_{\sigma}^{i}\phi^{\tau}(L\gamma_{m}L)(\gamma^{m}\lambda)_{\alpha}+\tfrac{1}{8\mathcal{T}}\delta_{\sigma}^{i}\varphi^{+}(P_{n}+\mathcal{T}\partial_{\sigma}X_{n})(L\gamma^{m}\gamma^{n}\Lambda)\tfrac{(\gamma^{m}\lambda)_{\alpha}}{(\Lambda\lambda)}\nonumber \\
 &  & +\tfrac{1}{4\mathcal{T}}\delta_{\tau}^{i}\varphi^{+}\big\{(P_{m}+\mathcal{T}\partial_{\sigma}X_{m})(P^{m}+\mathcal{T}\partial_{\sigma}X^{m})\big\}\tfrac{\Lambda_{\alpha}}{(\Lambda\lambda)},\nonumber \\
 &  & -\tfrac{1}{8\mathcal{T}}\delta_{\sigma}^{i}\varphi^{+}\big\{(P_{n}+\mathcal{T}\partial_{\sigma}X_{n})(\gamma^{n}\lambda)_{\beta}\big\}(\gamma_{m}L)_{\alpha}\tfrac{(\gamma^{m}\Lambda)^{\beta}}{(\Lambda\lambda)},\\
\delta^{'}[\delta A_{i}] & = & \phi^{j}\{\partial_{j}A_{i}-\partial_{i}A_{j}\},\\
\delta^{'}[\delta B] & = & \phi^{k}\epsilon_{ki}\{\lambda^{\alpha}w_{\alpha}^{i}+\epsilon^{ij}(\Sigma\chi_{j}+\partial_{j}B)\},\\
\delta^{'}[\delta L^{\alpha}] & = & \varphi^{+}(P_{m}+\mathcal{T}\partial_{\sigma}X_{m})\{\nabla_{\tau}\lambda^{\beta}\}\Lambda_{\beta}\tfrac{(\gamma^{m}\Lambda)^{\alpha}}{(\Lambda\lambda)^{2}}-\tfrac{1}{2}\varphi^{+}(L\gamma_{m})_{\beta}\{\nabla_{\sigma}\lambda^{\beta}\}\tfrac{(\gamma^{m}\Lambda)^{\alpha}}{(\Lambda\lambda)}\nonumber \\
 &  & -\varphi^{+}\Big\{\partial_{\tau}P_{m}-\tfrac{1}{4}\partial_{\sigma}(L\gamma_{m}\lambda)+\tfrac{1}{4}\partial_{\sigma}(\hat{L}\gamma_{m}\hat{\lambda})\Big\}\tfrac{(\gamma^{m}\Lambda)^{\alpha}}{(\Lambda\lambda)}\nonumber \\
 &  & -\mathcal{T}\varphi^{+}\partial_{\sigma}\Big\{\partial_{\tau}X_{m}-\tfrac{1}{4\mathcal{T}}(L\gamma^{m}\lambda)-\tfrac{1}{4\mathcal{T}}(\hat{L}\gamma^{m}\hat{\lambda})\Big\}\tfrac{(\gamma^{m}\Lambda)^{\alpha}}{(\Lambda\lambda)}\nonumber \\
 &  & -\tfrac{1}{8}\varphi^{+}(\Lambda\gamma_{mn}\nabla_{\sigma}L)\tfrac{(\gamma^{mn}\lambda)^{\alpha}}{(\Lambda\lambda)}-\tfrac{1}{4}\varphi^{+}(\Lambda\nabla_{\sigma}L)\tfrac{\lambda^{\alpha}}{(\Lambda\lambda)},\\
\delta^{'}[\delta\xi^{\alpha}] & = & \phi^{i}\{\partial_{i}\xi^{\alpha}-\chi_{i}\lambda^{\alpha}\},\\
\delta^{'}[\delta p_{\alpha}^{i}] & = & \phi^{i}\{\partial_{j}p_{\alpha}^{j}\},\\
\delta^{'}[\delta\chi_{i}] & = & \phi^{j}\{\nabla_{j}\chi_{i}-\nabla_{i}\chi_{j}\},\\
\delta^{'}[\delta\Sigma] & = & \phi^{k}\epsilon_{kj}\{\lambda^{\alpha}p_{\alpha}^{i}+\epsilon^{ij}\nabla_{j}\Sigma\}.
\end{eqnarray}
\end{subequations}Similar equations hold for the hatted sector. This
confirms that worldsheet reparametrization is a redundant symmetry
of the action, since the parameters $\phi^{i}$ and $\varphi^{\pm}$
can be used to set $c^{i}=a^{\pm}=0$. Furthermore, their gauge-for-gauge
transformations in \eqref{eq:g4g} involve only simple field shifts,
\emph{i.e.} no derivatives of the gauge-for-gauge parameters, therefore
generating no dynamical ghost-for-ghosts. Consequently, the gauge
symmetries parametrized by $c^{i}$ and $a^{\pm}$ can be disregarded
in the construction of the master action within the Batalin-Vilkovisky
formalism. It will be demonstrated next that the quantization of the
action $S_{0}$ leads to the pure spinor superstring.

\subsection{The pure spinor master action\label{subsec:PSmasteraction}}

In order to build the pure spinor master action in the Batalin-Vilkovisky
formalism, the gauge parameters discussed above will be promoted to
dynamical variables. The field content of the model will be collectively
denoted by $\Phi^{I}$, with the index $I$ running over the set
\begin{multline}
\Phi^{I}=\{X^{m},P_{m},w_{\alpha}^{i},\lambda^{\alpha},L^{\alpha},A_{i},B,p_{\alpha}^{i},\xi^{\alpha},\chi_{i},\Sigma,\Omega,\theta^{\alpha},s_{\alpha},\epsilon_{\alpha},\gamma,\\
\hat{w}_{\hat{\alpha}}^{i},\hat{\lambda}^{\hat{\alpha}},\hat{L}^{\hat{\alpha}},\hat{A}_{i},\hat{B},\hat{p}_{\hat{\alpha}}^{i},\hat{\xi}^{\hat{\alpha}},\hat{\chi}_{i},\hat{\Sigma},\hat{\Omega},\hat{\theta}^{\hat{\alpha}},\hat{s}_{\hat{\alpha}},\hat{\epsilon}_{\hat{\alpha}},\hat{\gamma}\}.
\end{multline}
As usual, ghost fields and the correspondent gauge parameters have
opposite statistics, therefore $\{\Omega,\hat{\Omega},\theta^{\alpha},\hat{\theta}^{\hat{\alpha}},s_{\alpha},\hat{s}_{\hat{\alpha}}\}$
are Grassmann odd while $\{\epsilon_{\alpha},\hat{\epsilon}_{\hat{\alpha}},\gamma,\hat{\gamma}\}$
are Grassmann even fields. Following the discussion at the end of
the previous subsection, the gauge parameters $c^{i}$ and $a^{\pm}$,
and the gauge-for-gauge parameters $\phi^{i}$ and $\varphi^{\pm}$
will be ignored.

For every field $\Phi^{I}$ there is an antifield $\Phi_{I}^{*}$
associated, with opposite statistics, and the antifield set is given
by
\begin{multline}
\Phi_{I}^{*}=\{X_{m}^{*},P_{*}^{m},w_{i*}^{\alpha},\lambda_{\alpha}^{*},L_{\alpha}^{*},A_{*}^{i},B^{*},p_{i*}^{\alpha},\xi_{\alpha}^{*},\chi_{*}^{i},\Sigma^{*},\Omega^{*},\theta_{\alpha}^{*},s_{*}^{\alpha},\epsilon_{*}^{\alpha},\gamma^{*},\\
\hat{w}_{i*}^{\hat{\alpha}},\hat{\lambda}_{\hat{\alpha}}^{*},\hat{L}_{\hat{\alpha}}^{*},\hat{A}_{*}^{i},\hat{B}^{*},\hat{p}_{i*}^{\hat{\alpha}},\hat{\xi}_{\hat{\alpha}}^{*},\hat{\chi}_{*}^{i},\hat{\Sigma}^{*},\hat{\Omega}^{*},\hat{\theta}_{\hat{\alpha}}^{*},\hat{s}_{*}^{\hat{\alpha}},\hat{\epsilon}_{*}^{\hat{\alpha}},\hat{\gamma}^{*}\}.
\end{multline}

By definition, fields and antifields are conjugate to each other,
satisfying the antibracket relation
\begin{equation}
\{\Phi_{I}^{*},\Phi^{J}\}=\delta_{I}^{J}.\label{eq:FaFconjugates}
\end{equation}
In general, the antibrackets between two operators $\mathcal{O}_{1}$
and $\mathcal{O}_{2}$ are defined as
\begin{equation}
\{\mathcal{O}_{1},\mathcal{O}_{2}\}\equiv\sum_{I}\left\{ \mathcal{O}_{1}\left(\frac{\overleftarrow{\partial}}{\partial\Phi_{I}^{*}}\frac{\partial}{\partial\Phi^{I}}-\frac{\overleftarrow{\partial}}{\partial\Phi^{I}}\frac{\partial}{\partial\Phi_{I}^{*}}\right)\mathcal{O}_{2}\right\} ,\label{eq:antibrackets}
\end{equation}
from which equation \eqref{eq:FaFconjugates} follows. However, as
consequence of the constraints \eqref{eq:PSconstraint} and \eqref{eq:SPSconstraint},
not all the components of $\Phi^{I}$ and $\Phi_{I}^{*}$ are independent.
In fact, the pure spinor constraints are generalized to
\begin{equation}
\begin{array}{rclcrcl}
(\lambda\gamma^{m}\lambda) & = & 0, &  & (\lambda\gamma^{m}p_{i*})+(\xi\gamma^{m}w_{i*}) & = & 0,\\
(\lambda\gamma^{m}\xi) & = & 0, &  & \lambda^{\alpha}\theta_{\alpha}^{*}+w_{\tau*}^{\alpha}L_{\alpha}^{*} & = & 0,\\
(\lambda\gamma^{m}w_{i*}) & = & 0, &  & \xi^{\alpha}\theta_{\alpha}^{*}+p_{\tau*}^{\alpha}L_{\alpha}^{*} & = & 0,\\
\lambda^{\alpha}L_{\alpha}^{*} & = & 0, &  & (\lambda\gamma^{mn}\theta^{*})+(w_{\tau*}\gamma^{mn}L^{*}) & = & 0,\\
(\lambda\gamma^{mn}L^{*}) & = & 0, &  & (\lambda\gamma^{m}s_{*})-\tfrac{1}{2}\epsilon^{ij}(w_{i*}\gamma^{m}w_{j*}) & = & 0,\\
\xi^{\alpha}L_{\alpha}^{*} & = & 0, &  & (\lambda\gamma^{m}\epsilon_{*})+(\xi\gamma^{m}s_{*})-\epsilon^{ij}(p_{i*}\gamma^{m}w_{j*}) & = & 0,
\end{array}\label{eq:PSconstraintsBV}
\end{equation}
and analogous constraints on the hatted sector, and the antibracket
\eqref{eq:FaFconjugates} cannot be naively computed. For example,
\begin{equation}
\{\lambda_{\alpha}^{*},(\lambda\gamma^{m}\lambda)\}=2(\gamma^{m}\lambda)_{\alpha},
\end{equation}
which is not compatible with the constraint $(\lambda\gamma^{m}\lambda)=0$.
This contradiction arises because there is an intrinsic gauge freedom
implied by the constraints \eqref{eq:PSconstraintsBV}, which only
have vanishing antibrackets with operators invariant under the pure
spinor gauge transformations given by\begin{subequations}\label{eq:PSsymmetriesFULL}
\begin{eqnarray}
\delta\lambda_{\alpha}^{*} & = & b_{m}(\gamma^{m}\lambda)_{\alpha}+c_{m}(\gamma^{m}\xi)_{\alpha}-d_{m}^{i}(\gamma^{m}w_{i*})_{\alpha}-e_{m}^{i}(\gamma^{m}p_{i*})_{\alpha}-fL_{\alpha}^{*}\nonumber \\
 &  & +f_{mn}(\gamma^{mn}L^{*})_{\alpha}-\bar{f}\theta_{\alpha}^{*}+\bar{f}_{mn}(\gamma^{mn}\theta^{*})_{\alpha}-h_{m}(\gamma^{m}s_{*})_{\alpha}-\bar{h}_{m}(\gamma^{m}\epsilon_{*})_{\alpha},\\
\delta\xi_{\alpha}^{*} & = & c_{m}(\gamma^{m}\lambda)_{\alpha}+e_{m}^{i}(\gamma^{m}w_{i*})_{\alpha}+gL_{\alpha}^{*}-\bar{g}\theta_{\alpha}^{*}-\bar{h}_{m}(\gamma^{m}s_{*})_{\alpha},\\
\delta w_{\alpha}^{i} & = & d_{m}^{i}(\gamma^{m}\lambda)_{\alpha}+e_{m}^{i}(\gamma^{m}\xi)_{\alpha}-\delta_{\tau}^{i}\bar{f}L_{\alpha}^{*}+\delta_{\tau}^{i}\bar{f}_{mn}(\gamma^{mn}L^{*})_{\alpha}+\epsilon^{ij}h_{m}(\gamma^{m}w_{j*})_{\alpha}\nonumber \\
 &  & +\epsilon^{ij}\bar{h}_{m}(\gamma^{m}p_{j*})_{\alpha},\\
\delta p_{\alpha}^{i} & = & e_{m}^{i}(\gamma^{m}\lambda)_{\alpha}+\delta_{\tau}^{i}\bar{g}L_{\alpha}^{*}-\epsilon^{ij}\bar{h}_{m}(\gamma^{m}w_{j*})_{\alpha},\\
\delta L^{\alpha} & = & f\lambda^{\alpha}+f_{mn}(\gamma^{mn}\lambda)^{\alpha}+g\xi^{\alpha}+\bar{f}w_{\tau*}^{\alpha}+\bar{f}_{mn}(\gamma^{mn}w_{\tau*})^{\alpha}+\bar{g}p_{\tau*}^{\alpha},\\
\delta\theta^{\alpha} & = & \bar{f}\lambda^{\alpha}+\bar{f}_{mn}(\gamma^{mn}\lambda)^{\alpha}+\bar{g}\xi^{\alpha},\\
\delta s_{\alpha} & = & h_{m}(\gamma^{m}\lambda)_{\alpha}+\bar{h}_{m}(\gamma^{m}\xi)_{\alpha},\\
\delta\epsilon_{\alpha} & = & \bar{h}_{m}(\gamma^{m}\lambda)_{\alpha},
\end{eqnarray}
\end{subequations}where $b_{m}$, $c_{m}$, $d_{m}$, $e_{m}$, $f$,
$f_{mn}$, $g$, $h_{m}$, $\bar{f}$ , $\bar{f}_{mn}$, $\bar{g}$
and $\bar{h}_{m}$ are local parameters. Again, similar transformations
exist for the hatted sector.

The pure spinor master action has to be concomitantly determined with
the pure spinor constraints \eqref{eq:PSconstraintsBV} and symmetries
\eqref{eq:PSsymmetriesFULL}. It can be cast as
\begin{equation}
S=S_{0}+S_{1}+S_{2}+S_{3},\label{eq:PSMASTER}
\end{equation}
where $S_{0}$ is displayed in \eqref{eq:newaction} and\begin{subequations}
\begin{eqnarray}
S_{1} & = & \int d\tau d\sigma\{\tfrac{1}{4\mathcal{T}}(\lambda\gamma^{m}\theta)X_{m}^{*}-\tfrac{1}{4}(\lambda\gamma_{m}\theta)\partial_{\sigma}P_{*}^{m}-\tfrac{1}{4\mathcal{T}}(P_{m}+T\partial_{\sigma}X_{m})(\theta\gamma^{m}w_{\tau*})\}\nonumber \\
 &  & +\int d\tau d\sigma\{\tfrac{1}{8\mathcal{T}}(L\gamma_{m}\lambda)(\theta\gamma^{m}w_{\sigma*})+(\nabla_{\tau}\theta^{\alpha})L_{\alpha}^{*}+\epsilon^{ij}(\nabla_{j}s_{\alpha})w_{i*}^{\alpha}-\lambda^{\alpha}s_{\alpha}B^{*}\}\nonumber \\
 &  & +\int d\tau d\sigma\{\Omega\lambda^{\alpha}\lambda_{\alpha}^{*}-\Omega w_{\alpha}^{i}w_{i*}^{\alpha}-(\partial_{i}\Omega)A_{*}^{i}-\Omega L^{\alpha}L_{\alpha}^{*}-\Omega\chi_{i}\chi_{*}^{i}+\Omega\Sigma\Sigma^{*}\}\nonumber \\
 &  & +\int d\tau d\sigma\{\gamma p_{\alpha}^{i}w_{i*}^{\alpha}+\gamma\Sigma B^{*}+\gamma\lambda^{\alpha}\xi_{\alpha}^{*}+(\nabla_{i}\gamma)\chi_{*}^{i}\}\nonumber \\
 &  & +\int d\tau d\sigma\{\epsilon^{ij}\epsilon_{\alpha}\chi_{j}w_{i*}^{\alpha}+\epsilon^{ij}(\partial_{j}\epsilon_{\alpha})p_{i*}^{\alpha}-\lambda^{\alpha}\epsilon_{\alpha}\Sigma^{*}\}\nonumber \\
 &  & +\int d\tau d\sigma\{\tfrac{1}{4\mathcal{T}}(\hat{\lambda}\gamma^{m}\hat{\theta})X_{m}^{*}+\tfrac{1}{4}(\hat{\lambda}\gamma_{m}\hat{\theta})\partial_{\sigma}P_{*}^{m}-\tfrac{1}{4\mathcal{T}}(P_{m}-\mathcal{T}\partial_{\sigma}X_{m})(\hat{\theta}\gamma^{m}\hat{w}_{\tau*})\}\nonumber \\
 &  & +\int d\tau d\sigma\{-\tfrac{1}{8\mathcal{T}}(\hat{L}\gamma_{m}\hat{\lambda})(\hat{\theta}\gamma^{m}\hat{w}_{\sigma*})+(\hat{\nabla}_{\tau}\hat{\theta}^{\hat{\alpha}})\hat{L}_{\hat{\alpha}}^{*}+\epsilon^{ij}(\hat{\nabla}_{j}\hat{s}_{\hat{\alpha}})\hat{w}_{i*}^{\hat{\alpha}}-\hat{\lambda}^{\hat{\alpha}}\hat{s}_{\hat{\alpha}}\hat{B}^{*}\}\nonumber \\
 &  & +\int d\tau d\sigma\{\hat{\Omega}\hat{\lambda}^{\hat{\alpha}}\hat{\lambda}_{\hat{\alpha}}^{*}-\hat{\Omega}\hat{w}_{\hat{\alpha}}^{i}\hat{w}_{i*}^{\hat{\alpha}}-(\partial_{i}\hat{\Omega})\hat{A}_{*}^{i}-\hat{\Omega}\hat{L}^{\hat{\alpha}}\hat{L}_{\hat{\alpha}}^{*}-\hat{\Omega}\hat{\chi}_{i}\hat{\chi}_{*}^{i}+\hat{\Omega}\hat{\Sigma}\hat{\Sigma}^{*}\}\nonumber \\
 &  & +\int d\tau d\sigma\{\hat{\gamma}\hat{p}_{\hat{\alpha}}^{i}\hat{w}_{i*}^{\hat{\alpha}}+\hat{\gamma}\hat{\Sigma}\hat{B}^{*}+\hat{\gamma}\hat{\lambda}^{\hat{\alpha}}\hat{\xi}_{\hat{\alpha}}^{*}+(\hat{\nabla}_{i}\hat{\gamma})\hat{\chi}_{*}^{i}\}\nonumber \\
 &  & +\int d\tau d\sigma\{\epsilon^{ij}\hat{\epsilon}_{\hat{\alpha}}\hat{\chi}_{j}\hat{w}_{i*}^{\hat{\alpha}}+\epsilon^{ij}(\partial_{j}\hat{\epsilon}_{\hat{\alpha}})\hat{p}_{i*}^{\hat{\alpha}}-\hat{\lambda}^{\hat{\alpha}}\hat{\epsilon}_{\hat{\alpha}}\hat{\Sigma}^{*}\},\\
S_{2} & = & \int d\tau d\sigma\{-\Omega\theta^{\alpha}\theta_{\alpha}^{*}-\Omega s_{\alpha}s_{*}^{\alpha}-\Omega\gamma\gamma^{*}-\gamma\epsilon_{\alpha}s_{*}^{\alpha}\}\nonumber \\
 &  & +\int d\tau d\sigma\{-\hat{\Omega}\hat{\theta}^{\hat{\alpha}}\hat{\theta}_{\hat{\alpha}}^{*}-\hat{\Omega}\hat{s}_{\hat{\alpha}}\hat{s}_{*}^{\hat{\alpha}}-\hat{\Omega}\hat{\gamma}\hat{\gamma}^{*}-\hat{\gamma}\hat{\epsilon}_{\hat{\alpha}}\hat{s}_{*}^{\hat{\alpha}}\},\\
S_{3} & = & \int d\tau d\sigma\{\tfrac{1}{16\mathcal{T}}(\theta\gamma^{m}s_{*})(\lambda\gamma_{m}\theta)+\tfrac{1}{16\mathcal{T}}(w_{\tau*}\gamma^{m}\theta)(w_{\sigma*}\gamma_{m}\theta)\}\nonumber \\
 &  & -\int d\tau d\sigma\{\tfrac{1}{16\mathcal{T}}(\hat{\theta}\gamma^{m}\hat{s}_{*})(\hat{\lambda}\gamma_{m}\hat{\theta})+\tfrac{1}{16\mathcal{T}}(\hat{w}_{\tau*}\gamma^{m}\hat{\theta})(\hat{w}_{\sigma*}\gamma_{m}\hat{\theta})\}.
\end{eqnarray}
\end{subequations}$S_{1}$ is connected to the gauge transformations
of the action $S_{0}$, while $S_{2}$ represents the extension of
the gauge algebra to the ghost fields. The last piece, $S_{3}$, is
required in order for $S$ to satisfy the master equation
\begin{equation}
\{S,S\}=0.
\end{equation}

By construction, the master action is invariant under the BV transformations
defined as
\begin{equation}
\delta_{\text{\tiny{BV}}}\mathcal{O}\equiv\{S,\mathcal{O}\},
\end{equation}
for any operator $\mathcal{O}$. Naturally, the BV transformations
of the fundamental fields in the action $S_{0}$ have a similar structure
to their gauge transformations and are given by\begin{subequations}\label{eq:BVfields}
\begin{eqnarray}
\delta_{\text{\tiny{BV}}}X^{m} & = & \tfrac{1}{4\mathcal{T}}(\lambda\gamma^{m}\theta)+\tfrac{1}{4\mathcal{T}}(\hat{\lambda}\gamma^{m}\hat{\theta}),\\
\delta_{\text{\tiny{BV}}}P_{m} & = & \tfrac{1}{4}\nabla_{\sigma}(\lambda\gamma^{m}\theta)-\tfrac{1}{4}\hat{\nabla}_{\sigma}(\hat{\lambda}\gamma^{m}\hat{\theta}),\\
\delta_{\text{\tiny{BV}}}\lambda^{\alpha} & = & \Omega\lambda^{\alpha},\\
\delta_{\text{\tiny{BV}}}w_{\alpha}^{i} & = & -\Omega w_{\alpha}^{i}+\epsilon^{ij}(\nabla_{j}s_{\alpha}+\epsilon_{\alpha}\chi_{j})+\delta_{\sigma}^{i}\tfrac{1}{8\mathcal{T}}[(L_{+}\gamma_{m}\lambda)-\tfrac{1}{2}(w_{\tau*}\gamma_{m}\theta)](\gamma^{m}\theta)_{\alpha}\nonumber \\
 &  & +\gamma p_{\alpha}^{i}-\delta_{\tau}^{i}\tfrac{1}{4\mathcal{T}}[P_{m}+\mathcal{T}\partial_{\sigma}X_{m}+\tfrac{1}{4}(w_{\sigma*}\gamma_{m}\theta)](\gamma^{m}\theta)_{\alpha},\\
\delta_{\text{\tiny{BV}}}A_{i} & = & -\partial_{i}\Omega,\\
\delta_{\text{\tiny{BV}}}B & = & \gamma\Sigma-\lambda^{\alpha}s_{\alpha},\\
\delta_{\text{\tiny{BV}}}L^{\alpha} & = & \nabla_{\tau}\theta^{\alpha}-\Omega L^{\alpha},\\
\delta_{\text{\tiny{BV}}}\xi^{\alpha} & = & \gamma\lambda^{\alpha},\\
\delta_{\text{\tiny{BV}}}p_{\alpha}^{i} & = & \epsilon^{ij}\partial_{j}\epsilon_{\alpha},\\
\delta_{\text{\tiny{BV}}}\chi_{i} & = & \nabla_{i}\gamma-\Omega\chi_{i},\\
\delta_{\text{\tiny{BV}}}\Sigma & = & \Omega\Sigma-\lambda^{\alpha}\epsilon_{\alpha},\\
\delta_{\text{\tiny{BV}}}\hat{\lambda}^{\hat{\alpha}} & = & \hat{\Omega}\hat{\lambda}^{\hat{\alpha}},\\
\delta_{\text{\tiny{BV}}}\hat{w}_{\hat{\alpha}}^{i} & = & -\Omega\hat{w}_{\hat{\alpha}}^{i}+\epsilon^{ij}(\hat{\nabla}_{j}\hat{s}_{\alpha}+\hat{\epsilon}_{\alpha}\hat{\chi}_{j})-\delta_{\sigma}^{i}\tfrac{1}{8\mathcal{T}}[(\hat{L}\gamma_{m}\hat{\lambda})-\tfrac{1}{2}(\hat{w}_{\tau*}\gamma_{m}\hat{\theta})](\gamma^{m}\hat{\theta})_{\hat{\alpha}}\nonumber \\
 &  & +\hat{\gamma}\hat{p}_{\hat{\alpha}}^{i}-\delta_{\tau}^{i}\tfrac{1}{4\mathcal{T}}[P_{m}-\mathcal{T}\partial_{\sigma}X_{m}-\tfrac{1}{4}(\hat{w}_{\sigma*}\gamma_{m}\hat{\theta})](\gamma^{m}\hat{\theta})_{\hat{\alpha}},\\
\delta_{\text{\tiny{BV}}}\hat{A}_{i} & = & -\partial_{i}\hat{\Omega},\\
\delta_{\text{\tiny{BV}}}\hat{B} & = & \hat{\gamma}\hat{\Sigma}-\hat{\lambda}^{\hat{\alpha}}\hat{s}_{\hat{\alpha}},\\
\delta_{\text{\tiny{BV}}}\hat{L}^{\hat{\alpha}} & = & \hat{\nabla}_{\tau}\hat{\theta}^{\hat{\alpha}}-\hat{\Omega}\hat{L}^{\hat{\alpha}},\\
\delta_{\text{\tiny{BV}}}\hat{\xi}^{\hat{\alpha}} & = & \hat{\gamma}\hat{\lambda}^{\hat{\alpha}},\\
\delta_{\text{\tiny{BV}}}\hat{p}_{\hat{\alpha}}^{i} & = & \epsilon^{ij}\partial_{j}\hat{\epsilon}_{\hat{\alpha}},\\
\delta_{\text{\tiny{BV}}}\hat{\chi}_{i} & = & \hat{\nabla}_{i}\hat{\gamma}-\hat{\Omega}\hat{\chi}_{i},\\
\delta_{\text{\tiny{BV}}}\hat{\Sigma} & = & \hat{\Omega}\hat{\Sigma}-\hat{\lambda}^{\hat{\alpha}}\hat{\epsilon}_{\hat{\alpha}}.
\end{eqnarray}
\end{subequations}In general, the BV transformations are nilpotent.
However, due to the pure spinor constraints \eqref{eq:PSconstraintsBV},
the transformations above are nilpotent up to pure spinor gauge transformations.
For example,\begin{subequations}
\begin{eqnarray}
\delta_{\text{\tiny{BV}}}^{2}w_{\alpha}^{i} & = & \tfrac{1}{16\mathcal{T}}\epsilon^{ij}(\theta\gamma_{m}\nabla_{j}\theta)(\gamma^{m}\lambda)_{\alpha},\\
\delta_{\text{\tiny{BV}}}^{2}\hat{w}_{\hat{\alpha}}^{i} & = & -\tfrac{1}{16\mathcal{T}}\epsilon^{ij}(\hat{\theta}\gamma^{m}\hat{\nabla}_{j}\hat{\theta})(\gamma^{m}\hat{\lambda})_{\hat{\alpha}}.
\end{eqnarray}
\end{subequations}

With the promotion of gauge parameters to ghost fields, their BV transformations
are nontrivial and can be cast as\begin{subequations}\label{eq:BVghosts}
\begin{eqnarray}
\delta_{\text{\tiny{BV}}}\theta^{\alpha} & = & -\Omega\theta^{\alpha},\\
\delta_{\text{\tiny{BV}}}s_{\alpha} & = & -\Omega s_{\alpha}-\gamma\epsilon_{\alpha}-\tfrac{1}{16\mathcal{T}}(\lambda\gamma_{m}\theta)(\gamma^{m}\theta)_{\alpha},\\
\delta_{\text{\tiny{BV}}}\Omega & = & 0,\\
\delta_{\text{\tiny{BV}}}\gamma & = & -\Omega\gamma,\\
\delta_{\text{\tiny{BV}}}\epsilon_{\alpha} & = & 0,\\
\delta_{\text{\tiny{BV}}}\hat{\theta}^{\hat{\alpha}} & = & -\hat{\Omega}\hat{\theta}^{\hat{\alpha}},\\
\delta_{\text{\tiny{BV}}}\hat{s}_{\hat{\alpha}} & = & -\hat{\Omega}\hat{s}_{\hat{\alpha}}-\hat{\gamma}\hat{\epsilon}_{\hat{\alpha}}+\tfrac{1}{16\mathcal{T}}(\hat{\lambda}\gamma_{m}\hat{\theta})(\gamma^{m}\hat{\theta})_{\hat{\alpha}},\\
\delta_{\text{\tiny{BV}}}\hat{\Omega} & = & 0,\\
\delta_{\text{\tiny{BV}}}\hat{\gamma} & = & -\hat{\Omega}\hat{\gamma},\\
\delta_{\text{\tiny{BV}}}\hat{\epsilon}_{\hat{\alpha}} & = & 0.
\end{eqnarray}
\end{subequations}

For completeness, the BV transformations for the antifields are given
by\begin{subequations}\label{eq:BVantifields}
\begin{eqnarray}
\delta_{\text{\tiny{BV}}}X_{m}^{*} & = & \partial_{\tau}P_{m}-\tfrac{1}{4}\partial_{\sigma}[(L\gamma_{m}\lambda)+(\theta\gamma_{m}w_{\tau*})-(\hat{L}\gamma_{m}\hat{\lambda})-(\hat{\theta}\gamma_{m}\hat{w}_{\tau*})],\\
\delta_{\text{\tiny{BV}}}P_{*}^{m} & = & -\partial_{\tau}X^{m}+\tfrac{1}{4\mathcal{T}}[(L\gamma^{m}\lambda)+(\theta\gamma^{m}w_{\tau*})+(\hat{L}\gamma^{m}\hat{\lambda})+(\hat{\theta}\gamma^{m}\hat{w}_{\tau*})],\\
\delta_{\text{\tiny{BV}}}\lambda_{\alpha}^{*} & = & \nabla_{i}w_{\alpha}^{i}-\chi_{i}p_{\alpha}^{i}+s_{\alpha}B^{*}+\tfrac{1}{4\mathcal{T}}[P^{m}+\mathcal{T}\partial_{\sigma}X^{m}-\tfrac{1}{2}(\theta\gamma^{m}w_{\sigma*})](\gamma_{m}L)_{\alpha}\nonumber \\
 &  & -\Omega\lambda_{\alpha}^{*}-\gamma\xi_{\alpha}^{*}+\epsilon_{\alpha}\Sigma^{*}+\tfrac{1}{4\mathcal{T}}[\eta^{mn}X_{n}^{*}-\mathcal{T}\partial_{\sigma}P_{*}^{m}-\tfrac{1}{4}(\theta\gamma^{m}s_{*})](\gamma_{m}\theta)_{\alpha},\\
\delta_{\text{\tiny{BV}}}w_{i*}^{\alpha} & = & -\nabla_{i}\lambda^{\alpha}+\Omega w_{i*}^{\alpha},\\
\delta_{\text{\tiny{BV}}}A_{*}^{i} & = & -\lambda^{\alpha}w_{\alpha}^{i}-\epsilon^{ij}\partial_{j}B-\Sigma\epsilon^{ij}\chi_{j}-\epsilon^{ij}s_{\alpha}w_{j*}^{\alpha}+\delta_{\tau}^{i}\theta^{\alpha}L_{\alpha}^{*}+\gamma\chi_{*}^{i},\\
\delta_{\text{\tiny{BV}}}B^{*} & = & -\epsilon^{ij}\partial_{i}A_{j},\\
\delta_{\text{\tiny{BV}}}L_{\alpha}^{*} & = & \tfrac{1}{4\mathcal{T}}[P^{m}+\mathcal{T}\partial_{\sigma}X^{m}-\tfrac{1}{2}(\theta\gamma^{m}w_{\sigma*})](\gamma_{m}\lambda)_{\alpha}+\Omega L_{\alpha}^{*},\\
\delta_{\text{\tiny{BV}}}\xi_{\alpha}^{*} & = & \partial_{i}p_{\alpha}^{i},\\
\delta_{\text{\tiny{BV}}}p_{i*}^{\alpha} & = & \partial_{i}\xi^{\alpha}-\lambda^{\alpha}\chi_{i}+\gamma w_{i*}^{\alpha},\\
\delta_{\text{\tiny{BV}}}\chi_{*}^{i} & = & \lambda^{\alpha}p_{\alpha}^{i}+\epsilon^{ij}\nabla_{j}\Sigma+\Omega\chi_{*}^{i}-\epsilon^{ij}\epsilon_{\alpha}w_{j*}^{\alpha},\\
\delta_{\text{\tiny{BV}}}\Sigma^{*} & = & -\epsilon^{ij}\nabla_{i}\chi_{j}-\Omega\Sigma^{*}+\gamma B^{*},\\
\delta_{\text{\tiny{BV}}}\hat{\lambda}_{\hat{\alpha}}^{*} & = & \hat{\nabla}_{i}\hat{w}_{\hat{\alpha}}^{i}-\hat{\chi}_{i}\hat{p}_{\hat{\alpha}}^{i}+\hat{s}_{\hat{\alpha}}\hat{B}^{*}+\tfrac{1}{4\mathcal{T}}[P^{m}-\mathcal{T}\partial_{\sigma}X^{m}+\tfrac{1}{2}(\hat{\theta}\gamma^{m}\hat{w}_{\sigma*})](\gamma_{m}\hat{L})_{\hat{\alpha}}\nonumber \\
 &  & -\hat{\Omega}\hat{\lambda}_{\hat{\alpha}}^{*}-\hat{\gamma}\hat{\xi}_{\hat{\alpha}}^{*}+\hat{\epsilon}_{\hat{\alpha}}\hat{\Sigma}^{*}+\tfrac{1}{4\mathcal{T}}[\eta^{mn}X_{n}^{*}+\mathcal{T}\partial_{\sigma}P_{*}^{m}+\tfrac{1}{4}(\hat{\theta}\gamma^{m}\hat{s}_{*})](\gamma_{m}\hat{\theta})_{\hat{\alpha}},\\
\delta_{\text{\tiny{BV}}}\hat{w}_{i*}^{\hat{\alpha}} & = & -\hat{\nabla}_{i}\hat{\lambda}^{\hat{\alpha}}+\hat{\Omega}\hat{w}_{i*}^{\hat{\alpha}},\\
\delta_{\text{\tiny{BV}}}\hat{A}_{*}^{i} & = & -\hat{\lambda}^{\hat{\alpha}}\hat{w}_{\hat{\alpha}}^{i}-\epsilon^{ij}\partial_{j}\hat{B}-\hat{\Sigma}\epsilon^{ij}\hat{\chi}_{j}-\epsilon^{ij}\hat{s}_{\hat{\alpha}}\hat{w}_{j*}^{\hat{\alpha}}+\delta_{\tau}^{i}\hat{\theta}^{\hat{\alpha}}\hat{L}_{\hat{\alpha}}^{*}+\hat{\gamma}\hat{\chi}_{*}^{i},\\
\delta_{\text{\tiny{BV}}}\hat{B}^{*} & = & -\epsilon^{ij}\partial_{i}\hat{A}_{j},\\
\delta_{\text{\tiny{BV}}}\hat{L}_{\hat{\alpha}}^{*} & = & \tfrac{1}{4\mathcal{T}}[P^{m}-\mathcal{T}\partial_{\sigma}X^{m}+\tfrac{1}{2}(\hat{\theta}\gamma^{m}\hat{w}_{\sigma*})](\gamma_{m}\hat{\lambda})_{\hat{\alpha}}+\hat{\Omega}\hat{L}_{\hat{\alpha}}^{*},\\
\delta_{\text{\tiny{BV}}}\hat{\xi}_{\hat{\alpha}}^{*} & = & \partial_{i}\hat{p}_{\hat{\alpha}}^{i},\\
\delta_{\text{\tiny{BV}}}\hat{p}_{i*}^{\hat{\alpha}} & = & \partial_{i}\hat{\xi}^{\hat{\alpha}}-\hat{\lambda}^{\hat{\alpha}}\hat{\chi}_{i}+\hat{\gamma}\hat{w}_{i*}^{\hat{\alpha}},\\
\delta_{\text{\tiny{BV}}}\hat{\chi}_{*}^{i} & = & \hat{\lambda}^{\hat{\alpha}}\hat{p}_{\hat{\alpha}}^{i}+\epsilon^{ij}\hat{\nabla}_{j}\hat{\Sigma}+\hat{\Omega}\hat{\chi}_{*}^{i}-\epsilon^{ij}\hat{\epsilon}_{\hat{\alpha}}\hat{w}_{j*}^{\hat{\alpha}},\\
\delta_{\text{\tiny{BV}}}\hat{\Sigma}^{*} & = & -\epsilon^{ij}\hat{\nabla}_{i}\hat{\chi}_{j}-\hat{\Omega}\hat{\Sigma}^{*}+\hat{\gamma}\hat{B}^{*},
\end{eqnarray}
\end{subequations}and for the ghost antifields,\begin{subequations}\label{eq:BVghostsantifields}
\begin{eqnarray}
\delta_{\text{\tiny{BV}}}\theta_{\alpha}^{*} & = & \tfrac{1}{4\mathcal{T}}[\eta^{mn}X_{n}^{*}-\mathcal{T}\partial_{\sigma}P_{*}^{m}-\tfrac{1}{4}(\theta\gamma^{m}s_{*})](\gamma_{m}\lambda)_{\alpha}+\Omega\theta_{\alpha}^{*}-\nabla_{\tau}L_{\alpha}^{*}\nonumber \\
 &  & +\tfrac{1}{8\mathcal{T}}[(L\gamma_{m}\lambda)-\tfrac{1}{2}(w_{\tau*}\gamma_{m}\theta)](\gamma^{m}w_{\sigma*})_{\alpha}+\tfrac{1}{16\mathcal{T}}(\lambda\gamma_{m}\theta)(\gamma^{m}s_{*})_{\alpha}\nonumber \\
 &  & -\tfrac{1}{4\mathcal{T}}[P_{m}+\mathcal{T}\partial_{\sigma}X_{m}+\tfrac{1}{4}(w_{\sigma*}\gamma_{m}\theta)](\gamma^{m}w_{\tau*})_{\alpha},\\
\delta_{\text{\tiny{BV}}}s_{*}^{\alpha} & = & \epsilon^{ij}\nabla_{i}w_{j*}^{\alpha}-\lambda^{\alpha}B^{*}+\Omega s_{*}^{\alpha},\\
\delta_{\text{\tiny{BV}}}\Omega^{*} & = & \lambda^{\alpha}\lambda_{\alpha}^{*}-w_{\alpha}^{i}w_{i*}^{\alpha}+\partial_{i}A_{*}^{i}-L^{\alpha}L_{\alpha}^{*}-\chi_{i}\chi_{*}^{i}+\Sigma\Sigma^{*}-\theta^{\alpha}\theta_{\alpha}^{*}-s_{\alpha}s_{*}^{\alpha}-\gamma\gamma^{*},\\
\delta_{\text{\tiny{BV}}}\gamma^{*} & = & -p_{\alpha}^{i}w_{i*}^{\alpha}-\Sigma B^{*}-\lambda^{\alpha}\xi_{\alpha}^{*}+\nabla_{i}\chi_{*}^{i}+\Omega\gamma^{*}+\epsilon_{\alpha}s_{*}^{\alpha},\\
\delta_{\text{\tiny{BV}}}\epsilon_{*}^{\alpha} & = & \epsilon^{ij}\chi_{i}w_{j*}^{\alpha}-\epsilon^{ij}\partial_{i}p_{j*}^{\alpha}+\lambda^{\alpha}\Sigma^{*}+\gamma s_{*}^{\alpha},\\
\delta_{\text{\tiny{BV}}}\hat{\theta}_{\hat{\alpha}}^{*} & = & \tfrac{1}{4\mathcal{T}}[\eta^{mn}X_{n}^{*}+\mathcal{T}\partial_{\sigma}P_{*}^{m}+\tfrac{1}{4}(\hat{\theta}\gamma^{m}\hat{s}_{*})](\gamma_{m}\hat{\lambda})_{\hat{\alpha}}+\hat{\Omega}\hat{\theta}_{\hat{\alpha}}^{*}-\hat{\nabla}_{\tau}\hat{L}_{\hat{\alpha}}^{*}\nonumber \\
 &  & -\tfrac{1}{8\mathcal{T}}[(\hat{L}\gamma_{m}\hat{\lambda})-\tfrac{1}{2}(\hat{w}_{\tau*}\gamma_{m}\hat{\theta})](\gamma^{m}\hat{w}_{\sigma*})_{\hat{\alpha}}-\tfrac{1}{16\mathcal{T}}(\hat{\lambda}\gamma_{m}\hat{\theta})(\gamma^{m}\hat{s}_{*})_{\hat{\alpha}}\nonumber \\
 &  & -\tfrac{1}{4\mathcal{T}}[P_{m}-\mathcal{T}\partial_{\sigma}X_{m}-\tfrac{1}{4}(\hat{w}_{\sigma*}\gamma_{m}\hat{\theta})](\gamma^{m}\hat{w}_{\tau*})_{\hat{\alpha}},\\
\delta_{\text{\tiny{BV}}}\hat{s}_{*}^{\hat{\alpha}} & = & \epsilon^{ij}\hat{\nabla}_{i}\hat{w}_{j*}^{\hat{\alpha}}-\hat{\lambda}^{\hat{\alpha}}\hat{B}^{*}+\hat{\Omega}\hat{s}_{*}^{\hat{\alpha}},\\
\delta_{\text{\tiny{BV}}}\hat{\Omega}^{*} & = & \hat{\lambda}^{\hat{\alpha}}\hat{\lambda}_{\hat{\alpha}}^{*}-\hat{w}_{\hat{\alpha}}^{i}\hat{w}_{i*}^{\hat{\alpha}}+\partial_{i}\hat{A}_{*}^{i}-\hat{L}^{\hat{\alpha}}\hat{L}_{\hat{\alpha}}^{*}-\hat{\chi}_{i}\hat{\chi}_{*}^{i}+\hat{\Sigma}\hat{\Sigma}^{*}-\hat{\theta}^{\hat{\alpha}}\hat{\theta}_{\hat{\alpha}}^{*}-\hat{s}_{\hat{\alpha}}\hat{s}_{*}^{\hat{\alpha}}-\hat{\gamma}\hat{\gamma}^{*},\\
\delta_{\text{\tiny{BV}}}\hat{\gamma}^{*} & = & -\hat{p}_{\hat{\alpha}}^{i}\hat{w}_{i*}^{\hat{\alpha}}-\hat{\Sigma}\hat{B}^{*}-\hat{\lambda}^{\hat{\alpha}}\hat{\xi}_{\hat{\alpha}}^{*}+\hat{\nabla}_{i}\hat{\chi}_{*}^{i}+\hat{\Omega}\hat{\gamma}^{*}+\hat{\epsilon}_{\hat{\alpha}}\hat{s}_{*}^{\hat{\alpha}},\\
\delta_{\text{\tiny{BV}}}\hat{\epsilon}_{*}^{\hat{\alpha}} & = & \epsilon^{ij}\hat{\chi}_{i}\hat{w}_{j*}^{\hat{\alpha}}-\epsilon^{ij}\partial_{i}\hat{p}_{j*}^{\hat{\alpha}}+\hat{\lambda}^{\hat{\alpha}}\hat{\Sigma}^{*}+\hat{\gamma}\hat{s}_{*}^{\hat{\alpha}}.
\end{eqnarray}
\end{subequations}

\subsection{Gauge fixing\label{subsec:Gauge-fixing}}

The procedure for gauge fixing in the Batalin-Vilkovisky formalism
is very straightforward. But before moving on, it is useful to discuss
the particular gauge to be chosen here.

Using the scaling transformations \eqref{eq:scalingSYM}, it is possible
to choose $A_{\tau}=\hat{A}_{\tau}=0$. Similarly, the fermionic gauge
transformations \eqref{eq:fermionSYM} can be used to partially fix
their gauge fields to $\chi_{\tau}=\hat{\chi}_{\tau}=0$.

For the curl symmetry \eqref{eq:curlSYM}, the choice will be $w_{\alpha}^{\sigma}=p_{\alpha}^{\sigma}=0$
and $\hat{w}_{\hat{\alpha}}^{\sigma}=\hat{p}_{\hat{\alpha}}^{\sigma}=0$.
Observe here that there is no residual gauge transformation, since
both fields ($w_{\alpha}^{\sigma}$, $p_{\alpha}^{\sigma}$, $\hat{w}_{\hat{\alpha}}^{\sigma}$
and $\hat{p}_{\hat{\alpha}}^{\sigma}$) and ghosts ($s_{\alpha}$,
$\epsilon_{\alpha}$, $\hat{s}_{\hat{\alpha}}$ and $\hat{\epsilon}_{\hat{\alpha}}$)
have the same number of independent components according to the pure
spinor constraints \eqref{eq:PSconstraintsBV} and gauge transformations
\eqref{eq:PSsymmetriesFULL}.

Finally, the twistor-like symmetry \eqref{eq:twistorSYM} can be used
to fix the Lagrange multipliers $L^{\alpha}$ and $\hat{L}^{\hat{\alpha}}$.
Following the discussion after equation \eqref{eq:Berkovitsactioncovariant},
the gauge $L^{\alpha}=\hat{L}^{\hat{\alpha}}=0$ would imply a degenerate
worldsheet metric. Although worldsheet reparametrization is hidden
in the twistor-like symmetry, the conformal gauge would still be the
natural choice. In the Polyakov action \eqref{eq:Pol1st}, this gauge
would be $e_{\pm}=1$. For the action $S_{0}$ in \eqref{eq:newaction},
the conformal gauge is equivalent to the choice\begin{subequations}\label{eq:conformalgauge}
\begin{eqnarray}
L^{\alpha} & = & \tfrac{(\gamma^{m}\Lambda)^{\alpha}}{(\Lambda\lambda)}(P_{m}+\mathcal{T}\partial_{\sigma}X_{m}),\\
\hat{L}^{\hat{\alpha}} & = & \tfrac{(\gamma^{m}\hat{\Lambda})^{\hat{\alpha}}}{(\hat{\Lambda}\hat{\lambda})}(P_{m}-\mathcal{T}\partial_{\sigma}X_{m}).
\end{eqnarray}
\end{subequations}

The proposed gauge can be implemented through the gauge fixing fermion
$\Xi$, given by
\begin{multline}
\Xi=\int d\tau d\sigma\{\bar{\Omega}A_{\tau}+\beta\chi_{\tau}-r^{\alpha}w_{\alpha}^{\sigma}+\eta^{\alpha}p_{\alpha}^{\sigma}+\hat{\bar{\Omega}}\hat{A}_{\tau}-\hat{r}^{\hat{\alpha}}\hat{w}_{\hat{\alpha}}^{\sigma}+\hat{\beta}\hat{\chi}_{\tau}+\hat{\eta}^{\hat{\alpha}}\hat{p}_{\hat{\alpha}}^{\sigma}\}\\
-\int d\tau d\sigma\{\pi_{\alpha}[L^{\alpha}-\tfrac{(\gamma^{m}\Lambda)^{\alpha}}{(\Lambda\lambda)}(P_{m}+\mathcal{T}\partial_{\sigma}X_{m})]+\hat{\pi}_{\hat{\alpha}}[\hat{L}^{\hat{\alpha}}-\tfrac{(\gamma^{m}\hat{\Lambda})^{\hat{\alpha}}}{(\hat{\Lambda}\hat{\lambda})}(P_{m}-\mathcal{T}\partial_{\sigma}X_{m})]\}.\label{eq:GFF}
\end{multline}
Here, $\bar{\Omega}$, $\beta$, $r^{\alpha}$, $\eta^{\alpha}$ and
$\pi_{\alpha}$ are the antighosts of $\Omega$, $\gamma$, $s_{\alpha}$,
$\epsilon_{\alpha}$ and $\theta^{\alpha}$, respectively (hatted
and unhatted). The antighosts will be generically represented by $\bar{\Phi}_{a}$
(with antifields $\bar{\Phi}_{*}^{a}$), where the index $a$ denotes
the different components of the gauge parameters. In addition, the
extended phase space of the model will include the Nakanishi-Lautrup
fields, $\Lambda_{a}$, and antifields, $\Lambda_{*}^{a}$.

The gauge fixed action can be determined by evaluating the non-minimal
master action, defined as
\begin{equation}
S_{nm}=S+\int d\tau d\sigma\left(\bar{\Phi}_{*}^{a}\Lambda_{a}\right),
\end{equation}
at
\begin{equation}
\begin{array}{rclcrclcrcl}
\Phi_{I}^{*} & = & \frac{\delta\Xi}{\delta\Phi^{I}}, &  & \bar{\Phi}_{*}^{a} & = & \frac{\delta\Xi}{\delta\bar{\Phi}_{a}}, &  & \Lambda_{*}^{a} & = & \frac{\delta\Xi}{\delta\Lambda_{a}}.\end{array}
\end{equation}
For the particular choice of gauge fixing fermion \eqref{eq:GFF},
the non-vanishing antifields are given by 
\begin{equation}
\begin{array}{rclcrcl}
X_{m}^{*} & = & -\mathcal{T}\partial_{\sigma}[\tfrac{(\pi\gamma^{m}\Lambda)}{(\Lambda\lambda)}]+\mathcal{T}\partial_{\sigma}[\tfrac{(\hat{\pi}\gamma^{m}\hat{\Lambda})}{(\hat{\Lambda}\hat{\lambda})}], &  & A_{*}^{\tau} & = & \bar{\Omega},\\
\delta P_{*}^{m} & = & \mathcal{T}\tfrac{(\pi\gamma^{m}\Lambda)}{(\Lambda\lambda)}+\mathcal{T}\tfrac{(\hat{\pi}\gamma^{m}\hat{\Lambda})}{(\hat{\Lambda}\hat{\lambda})}, &  & \hat{A}_{*}^{\tau} & = & \hat{\bar{\Omega}},\\
\lambda_{\alpha}^{*} & = & -\tfrac{(\pi\gamma^{m}\Lambda)}{(\Lambda\lambda)^{2}}(P_{m}+\mathcal{T}\partial_{\sigma}X_{m})\Lambda_{\alpha}, &  & \chi_{*}^{\tau} & = & \beta,\\
\hat{\lambda}_{\hat{\alpha}}^{*} & = & -\tfrac{(\hat{\pi}\gamma^{m}\hat{\Lambda})}{(\hat{\Lambda}\hat{\lambda})^{2}}(P_{m}-\mathcal{T}\partial_{\sigma}X_{m})\hat{\Lambda}_{\hat{\alpha}}, &  & \hat{\chi}_{*}^{\tau} & = & \hat{\beta},\\
L_{\alpha}^{*} & = & -\pi_{\alpha}, &  & \hat{L}_{\hat{\alpha}}^{*} & = & -\hat{\pi}_{\hat{\alpha}},\\
w_{\sigma*}^{\alpha} & = & -r^{\alpha}, &  & \hat{w}_{\sigma*}^{\hat{\alpha}} & = & -\hat{r}^{\hat{\alpha}},\\
p_{\sigma*}^{\alpha} & = & \eta^{\alpha}, &  & \hat{p}_{\sigma*}^{\hat{\alpha}} & = & \hat{\eta}^{\hat{\alpha}}.
\end{array}
\end{equation}

Since the master action $S$ was consistently built with the pure
spinor constraints \eqref{eq:PSconstraintsBV}, the antighosts are
constrained accordingly:
\begin{equation}
\begin{array}{rclcrcl}
\xi^{\alpha}\pi_{\alpha} & = & 0, &  & \hat{\xi}^{\hat{\alpha}}\hat{\pi}_{\hat{\alpha}} & = & 0,\\
\lambda^{\alpha}\pi_{\alpha} & = & 0, &  & \hat{\lambda}^{\hat{\alpha}}\hat{\pi}_{\hat{\alpha}} & = & 0,\\
(\lambda\gamma^{m}r) & = & 0, &  & (\hat{\lambda}\gamma^{m}\hat{r}) & = & 0,\\
(\lambda\gamma^{mn}\pi) & = & 0, &  & (\hat{\lambda}\gamma^{mn}\hat{\pi}) & = & 0,\\
(\lambda\gamma^{m}\eta)+(r\gamma^{m}\xi) & = & 0, &  & (\hat{\lambda}\gamma^{m}\hat{\eta})+(\hat{r}\gamma^{m}\hat{\xi}) & = & 0.
\end{array}\label{eq:PSconstraintsantighosts}
\end{equation}

After solving for the equations of motion of the Nakanishi-Lautrup
fields, the gauge fixed action can be written as
\begin{eqnarray}
S_{fixed} & = & \int d\tau d\sigma\{P_{m}\partial_{\tau}X^{m}-\tfrac{1}{2\mathcal{T}}(P_{m}P^{m}+\mathcal{T}^{2}\partial_{\sigma}X_{m}\partial_{\sigma}X^{m})\}\nonumber \\
 &  & +\int d\tau d\sigma\{w_{\alpha}\partial_{\tau}\lambda^{\alpha}+p_{\alpha}\partial_{\tau}\xi^{\alpha}+\pi_{\alpha}\partial_{\tau}\theta^{\alpha}+r^{\alpha}\partial_{\tau}s_{\alpha}+\eta^{\alpha}\partial_{\tau}\epsilon_{\alpha}\}\nonumber \\
 &  & +\int d\tau d\sigma\{A\partial_{\tau}B+\Sigma\partial_{\tau}\chi+\bar{\Omega}\partial_{\tau}\Omega+\beta\partial_{\tau}\gamma\}\nonumber \\
 &  & +\int d\tau d\sigma\{\tfrac{1}{2}\partial_{\sigma}(\lambda\gamma^{m}\theta)\tfrac{(\pi\gamma^{m}\Lambda)}{(\Lambda\lambda)}-\tfrac{1}{8T}\tfrac{(\lambda\gamma^{n}\gamma^{m}\Lambda)}{(\Lambda\lambda)}(P_{m}+\mathcal{T}\partial_{\sigma}X_{m})(\theta\gamma_{n}r)\}\nonumber \\
 &  & +\int d\tau d\sigma\{\hat{w}_{\hat{\alpha}}\partial_{\tau}\hat{\lambda}^{\hat{\alpha}}+\hat{p}_{\hat{\alpha}}\partial_{\tau}\hat{\xi}^{\hat{\alpha}}+\hat{\pi}_{\hat{\alpha}}\partial_{\tau}\hat{\theta}^{\hat{\alpha}}+\hat{r}^{\hat{\alpha}}\partial_{\tau}\hat{s}_{\hat{\alpha}}+\hat{\eta}^{\hat{\alpha}}\partial_{\tau}\hat{\epsilon}_{\hat{\alpha}}\}\nonumber \\
 &  & +\int d\tau d\sigma\{\hat{A}\partial_{\tau}\hat{B}+\hat{\Sigma}\partial_{\tau}\hat{\chi}+\hat{\bar{\Omega}}\partial_{\tau}\hat{\Omega}+\hat{\beta}\partial_{\tau}\hat{\gamma}\}\nonumber \\
 &  & +\int d\tau d\sigma\{-\tfrac{1}{2}\partial_{\sigma}(\hat{\lambda}\gamma^{m}\hat{\theta})\tfrac{(\hat{\pi}\gamma^{m}\hat{\Lambda})}{(\hat{\Lambda}\hat{\lambda})}+\tfrac{1}{8T}\tfrac{(\hat{\lambda}\gamma^{n}\gamma^{m}\hat{\Lambda})}{(\hat{\Lambda}\hat{\lambda})}(P_{m}-\mathcal{T}\partial_{\sigma}X_{m})(\hat{\theta}\gamma_{n}\hat{r})\},\label{eq:PSfixed1}
\end{eqnarray}
where some of the fields were renamed in order to simplify the notation,
\begin{equation}
\begin{array}{rclcrclcrclcrcl}
w_{\alpha} & \equiv & w_{\alpha}^{\tau}, &  & A & \equiv & A_{\sigma}, &  & \hat{w}_{\hat{\alpha}} & \equiv & \hat{w}_{\hat{\alpha}}^{\tau}, &  & \hat{A} & \equiv & \hat{A}_{\sigma},\\
p_{\alpha} & \equiv & p_{\alpha}^{\tau}, &  & \chi & \equiv & \chi_{\sigma}, &  & \hat{p}_{\hat{\alpha}} & \equiv & \hat{p}_{\hat{\alpha}}^{\tau}, &  & \hat{\chi} & \equiv & \hat{\chi}_{\sigma}.
\end{array}\label{eq:rename}
\end{equation}

The BV-BRST transformations can be readily determined next. From equation
\eqref{eq:BVfields}, the transformations of the fields can be written
as\begin{subequations}\label{eq:BV-BRSTfields}
\begin{eqnarray}
\delta X^{m} & = & \tfrac{1}{4\mathcal{T}}[(\lambda\gamma^{m}\theta)+(\hat{\lambda}\gamma^{m}\hat{\theta})],\\
\delta P_{m} & = & \tfrac{1}{4}\partial_{\sigma}[(\lambda\gamma_{m}\theta)-(\hat{\lambda}\gamma_{m}\hat{\theta})],\\
\delta\lambda^{\alpha} & = & \Omega\lambda^{\alpha},\\
\delta w_{\alpha} & = & \gamma p_{\alpha}-\epsilon_{\alpha}\chi-\Omega w_{\alpha}-\nabla_{\sigma}s_{\alpha}-\tfrac{1}{4\mathcal{T}}[P_{m}+\mathcal{T}\partial_{\sigma}X_{m}-\tfrac{1}{4}(r\gamma_{m}\theta)](\theta\gamma^{m})_{\alpha},\\
\delta A & = & -\partial_{\sigma}\Omega,\\
\delta B & = & \gamma\Sigma-\lambda^{\alpha}s_{\alpha},\\
\delta\xi^{\alpha} & = & \gamma\lambda^{\alpha},\\
\delta p_{\alpha} & = & -\partial_{\sigma}\epsilon_{\alpha},\\
\delta\chi & = & \partial_{\sigma}\gamma-A\gamma-\Omega\chi,\\
\delta\Sigma & = & \Omega\Sigma-\lambda^{\alpha}\epsilon_{\alpha},\\
\delta\hat{\lambda}^{\hat{\alpha}} & = & \hat{\Omega}\hat{\lambda}^{\hat{\alpha}},\\
\delta\hat{w}_{\hat{\alpha}} & = & \hat{\gamma}\hat{p}_{\hat{\alpha}}-\hat{\epsilon}_{\hat{\alpha}}\hat{\chi}-\Omega\hat{w}_{\hat{\alpha}}-\hat{\nabla}_{\sigma}\hat{s}_{\hat{\alpha}}-\tfrac{1}{4\mathcal{T}}[P_{m}-\mathcal{T}\partial_{\sigma}X_{m}+\tfrac{1}{4}(\hat{r}\gamma_{m}\hat{\theta})](\gamma^{m}\hat{\theta})_{\hat{\alpha}},\\
\delta\hat{A} & = & -\partial_{\sigma}\hat{\Omega},\\
\delta\hat{B} & = & \hat{\gamma}\hat{\Sigma}-\hat{\lambda}^{\hat{\alpha}}\hat{s}_{\hat{\alpha}},\\
\delta\hat{\xi}^{\hat{\alpha}} & = & \hat{\gamma}\hat{\lambda}^{\hat{\alpha}},\\
\delta\hat{p}_{\hat{\alpha}} & = & -\partial_{\sigma}\hat{\epsilon}_{\hat{\alpha}},\\
\delta\hat{\chi} & = & \partial_{\sigma}\hat{\gamma}-\hat{A}\hat{\gamma}-\hat{\Omega}\hat{\chi},\\
\delta\hat{\Sigma} & = & \hat{\Omega}\hat{\Sigma}-\hat{\lambda}^{\hat{\alpha}}\hat{\epsilon}_{\hat{\alpha}},
\end{eqnarray}
\end{subequations}and, using equations \eqref{eq:BVghosts} and \eqref{eq:BVantifields},
the transformations of the ghosts are given by\begin{subequations}\label{eq:BV-BRSTghosts}
\begin{eqnarray}
\delta\pi_{\alpha} & = & \Omega\pi_{\alpha}-\tfrac{1}{4\mathcal{T}}(\gamma_{m}\lambda)_{\alpha}(P^{m}+\mathcal{T}\partial_{\sigma}X^{m})-\tfrac{1}{8\mathcal{T}}(\gamma_{m}\lambda)_{\alpha}(\theta\gamma^{m}r),\\
\delta\theta^{\alpha} & = & -\Omega\theta^{\alpha},\\
\delta r^{\alpha} & = & \partial_{\sigma}\lambda^{\alpha}+A\lambda^{\alpha}+\Omega r^{\alpha},\\
\delta s_{\alpha} & = & -\Omega s_{\alpha}-\gamma\epsilon_{\alpha}-\tfrac{1}{16\mathcal{T}}(\lambda\gamma_{m}\theta)(\gamma^{m}\theta)_{\alpha},\\
\delta\bar{\Omega} & = & -\lambda^{\alpha}w_{\alpha}+\partial_{\sigma}B+r^{\alpha}s_{\alpha}+\pi_{\alpha}\theta^{\alpha}+\Sigma\chi+\beta\gamma,\\
\delta\Omega & = & 0,\\
\delta\beta & = & \lambda^{\alpha}p_{\alpha}-\partial_{\sigma}\Sigma-A\Sigma+\Omega\beta-r^{\alpha}\epsilon_{\alpha},\\
\delta\gamma & = & -\Omega\gamma,\\
\delta\eta^{\alpha} & = & \partial_{\sigma}\xi^{\alpha}-\lambda^{\alpha}\chi-\gamma r^{\alpha},\\
\delta\epsilon_{\alpha} & = & 0,\\
\delta\hat{\pi}_{\hat{\alpha}} & = & \hat{\Omega}\hat{\pi}_{\hat{\alpha}}-\tfrac{1}{4\mathcal{T}}(\gamma_{m}\hat{\lambda})_{\hat{\alpha}}(P^{m}-\mathcal{T}\partial_{\sigma}X^{m})+\tfrac{1}{8\mathcal{T}}(\gamma_{m}\hat{\lambda})_{\hat{\alpha}}(\hat{\theta}\gamma^{m}\hat{r}),\\
\delta\hat{\theta}^{\hat{\alpha}} & = & -\hat{\Omega}\hat{\theta}^{\hat{\alpha}},\\
\delta\hat{r}^{\hat{\alpha}} & = & \partial_{\sigma}\hat{\lambda}^{\hat{\alpha}}+\hat{A}\hat{\lambda}^{\hat{\alpha}}+\hat{\Omega}r^{\hat{\alpha}},\\
\delta\hat{s}_{\hat{\alpha}} & = & -\hat{\Omega}\hat{s}_{\hat{\alpha}}-\hat{\gamma}\hat{\epsilon}_{\hat{\alpha}}+\tfrac{1}{16\mathcal{T}}(\hat{\lambda}\gamma_{m}\hat{\theta})(\gamma^{m}\hat{\theta})_{\hat{\alpha}},\\
\delta\hat{\bar{\Omega}} & = & -\hat{\lambda}^{\hat{\alpha}}\hat{w}_{\hat{\alpha}}+\partial_{\sigma}\hat{B}+\hat{r}^{\hat{\alpha}}\hat{s}_{\hat{\alpha}}+\hat{\pi}_{\hat{\alpha}}\hat{\theta}^{\hat{\alpha}}+\hat{\Sigma}\hat{\chi}+\hat{\beta}\hat{\gamma},\\
\delta\hat{\Omega} & = & 0,\\
\delta\hat{\beta} & = & \hat{\lambda}^{\hat{\alpha}}\hat{p}_{\hat{\alpha}}-\partial_{\sigma}\hat{\Sigma}-\hat{A}\hat{\Sigma}+\hat{\Omega}\hat{\beta}-\hat{r}^{\hat{\alpha}}\hat{\epsilon}_{\hat{\alpha}},\\
\delta\hat{\gamma} & = & -\hat{\Omega}\hat{\gamma},\\
\delta\hat{\eta}^{\hat{\alpha}} & = & \partial_{\sigma}\hat{\xi}^{\hat{\alpha}}-\hat{\lambda}^{\hat{\alpha}}\hat{\chi}-\hat{\gamma}\hat{r}^{\hat{\alpha}},\\
\delta\hat{\epsilon}_{\hat{\alpha}} & = & 0.
\end{eqnarray}
\end{subequations}

The action \eqref{eq:PSfixed1} does not look like an ordinary action
in the conformal gauge. This can be fixed with the addition of a BRST
trivial expression. Consider the operators\begin{subequations}
\begin{eqnarray}
I & \equiv & -\tfrac{1}{2}(r\gamma^{m}\theta)\tfrac{(\pi\gamma^{m}\Lambda)}{(\Lambda\lambda)}+w_{\alpha}r^{\alpha}-p_{\alpha}\eta^{\alpha}+A\bar{\Omega}+\chi\beta,\\
\hat{I} & \equiv & \tfrac{1}{2}(\hat{r}\gamma^{m}\hat{\theta})\tfrac{(\hat{\pi}\gamma^{m}\hat{\Lambda})}{(\hat{\Lambda}\hat{\lambda})}+\hat{w}_{\hat{\alpha}}\hat{r}^{\hat{\alpha}}-\hat{p}_{\hat{\alpha}}\hat{\eta}^{\hat{\alpha}}+\hat{A}\hat{\bar{\Omega}}+\hat{\chi}\hat{\beta},
\end{eqnarray}
\end{subequations}and their BRST variation. It is then straightforward
to demonstrate that
\begin{eqnarray}
\tilde{S} & \equiv & S_{fixed}+\left\{ Q,\int d\tau d\sigma\,(I+\hat{I})\right\} \nonumber \\
 & = & \int d\tau d\sigma\{P_{m}\partial_{\tau}X^{m}-\tfrac{1}{2\mathcal{T}}(P_{m}P^{m}+\mathcal{T}^{2}\partial_{\sigma}X_{m}\partial_{\sigma}X^{m})\}\nonumber \\
 &  & +\int d\tau d\sigma\{w_{\alpha}\partial_{-}\lambda^{\alpha}+p_{\alpha}\partial_{-}\xi^{\alpha}+\pi_{\alpha}\partial_{\tau}\theta^{\alpha}+r^{\alpha}\partial_{-}s_{\alpha}+\eta^{\alpha}\partial_{-}\epsilon_{\alpha}\}\nonumber \\
 &  & +\int d\tau d\sigma\{A\partial_{-}B+\Sigma\partial_{-}\chi+\bar{\Omega}\partial_{-}\Omega+\beta\partial_{-}\gamma+\tfrac{1}{2}(\lambda\gamma^{m}\partial_{\sigma}\theta)\tfrac{(\pi\gamma^{m}\Lambda)}{(\Lambda\lambda)}\}\nonumber \\
 &  & +\int d\tau d\sigma\{\hat{w}_{\hat{\alpha}}\partial_{+}\hat{\lambda}^{\hat{\alpha}}+\hat{p}_{\hat{\alpha}}\partial_{+}\hat{\xi}^{\hat{\alpha}}+\hat{\pi}_{\hat{\alpha}}\partial_{\tau}\hat{\theta}^{\hat{\alpha}}+\hat{r}^{\hat{\alpha}}\partial_{+}\hat{s}_{\hat{\alpha}}+\hat{\eta}^{\hat{\alpha}}\partial_{+}\hat{\epsilon}_{\hat{\alpha}}\}\nonumber \\
 &  & +\int d\tau d\sigma\{\hat{A}\partial_{+}\hat{B}+\hat{\Sigma}\partial_{+}\hat{\chi}+\hat{\bar{\Omega}}\partial_{+}\hat{\Omega}+\hat{\beta}\partial_{+}\hat{\gamma}-\tfrac{1}{2}(\hat{\lambda}\gamma^{m}\partial_{\sigma}\hat{\theta})\tfrac{(\hat{\pi}\gamma^{m}\hat{\Lambda})}{(\hat{\Lambda}\hat{\lambda})}\},\label{eq:PSfixed2}
\end{eqnarray}
where $\partial_{\pm}=\partial_{\tau}\pm\partial_{\sigma}$. Now,
using the gamma matrix identity
\begin{equation}
(\gamma^{mn})_{\alpha}^{\hphantom{\alpha}\beta}(\gamma_{mn})_{\lambda}^{\hphantom{\lambda}\gamma}=4\gamma_{\alpha\lambda}^{m}\gamma_{m}^{\beta\gamma}-2\delta_{\alpha}^{\beta}\delta_{\lambda}^{\gamma}-8\delta_{\lambda}^{\beta}\delta_{\alpha}^{\gamma},\label{eq:gammaproperty}
\end{equation}
it follows that\begin{subequations}
\begin{eqnarray}
-\tfrac{1}{2}(\lambda\gamma_{m}\partial_{\sigma}\theta)(\pi\gamma^{m}\Lambda) & = & \tfrac{1}{8}(\pi\gamma^{mn}\lambda)(\Lambda\gamma_{mn}\partial_{\sigma}\theta)+\tfrac{1}{4}(\pi\lambda)(\Lambda\partial_{\sigma}\theta)+(\Lambda\lambda)(\pi\partial_{\sigma}\theta),\\
-\tfrac{1}{2}(\hat{\lambda}\gamma_{m}\partial_{\sigma}\hat{\theta})(\hat{\pi}\gamma^{m}\hat{\Lambda}) & = & \tfrac{1}{8}(\hat{\pi}\gamma^{mn}\hat{\lambda})(\hat{\Lambda}\gamma_{mn}\partial_{\sigma}\hat{\theta})+\tfrac{1}{4}(\hat{\pi}\hat{\lambda})(\hat{\Lambda}\partial_{\sigma}\hat{\theta})+(\hat{\Lambda}\hat{\lambda})(\hat{\pi}\partial_{\sigma}\hat{\theta}).
\end{eqnarray}
\end{subequations}The first two terms on the right hand side of each
equation vanish due to the constraints \eqref{eq:PSconstraintsantighosts}.
Therefore the gauge fixed action \eqref{eq:PSfixed2} can be finally
rewritten as
\begin{eqnarray}
\tilde{S} & = & \int d\tau d\sigma\{P_{m}\partial_{\tau}X^{m}-\tfrac{1}{2\mathcal{T}}(P_{m}P^{m}+\mathcal{T}^{2}\partial_{\sigma}X_{m}\partial_{\sigma}X^{m})\}\nonumber \\
 &  & +\int d\tau d\sigma\{w_{\alpha}\partial_{-}\lambda^{\alpha}+p_{\alpha}\partial_{-}\xi^{\alpha}+\pi_{\alpha}\partial_{-}\theta^{\alpha}+r^{\alpha}\partial_{-}s_{\alpha}+\eta^{\alpha}\partial_{-}\epsilon_{\alpha}\}\nonumber \\
 &  & +\int d\tau d\sigma\{A\partial_{-}B+\Sigma\partial_{-}\chi+\bar{\Omega}\partial_{-}\Omega+\beta\partial_{-}\gamma\}\nonumber \\
 &  & +\int d\tau d\sigma\{\hat{w}_{\hat{\alpha}}\partial_{+}\hat{\lambda}^{\hat{\alpha}}+\hat{p}_{\hat{\alpha}}\partial_{+}\hat{\xi}^{\hat{\alpha}}+\hat{\pi}_{\hat{\alpha}}\partial_{+}\hat{\theta}^{\hat{\alpha}}+\hat{r}^{\hat{\alpha}}\partial_{+}\hat{s}_{\hat{\alpha}}+\hat{\eta}^{\hat{\alpha}}\partial_{+}\hat{\epsilon}_{\hat{\alpha}}\}\nonumber \\
 &  & +\int d\tau d\sigma\{\hat{A}\partial_{+}\hat{B}+\hat{\Sigma}\partial_{+}\hat{\chi}+\hat{\bar{\Omega}}\partial_{+}\hat{\Omega}+\hat{\beta}\partial_{+}\hat{\gamma}\},\label{eq:PSfixed3}
\end{eqnarray}
in which the separation between right and left-moving sectors is manifest.

The BRST current can be computed using the transformations \eqref{eq:BV-BRSTfields}
and \eqref{eq:BV-BRSTghosts}. It has two components, one left and
one right-moving, given by\begin{subequations}\label{eq:PS-BRSTcurrent}
\begin{eqnarray}
J & = & -\tfrac{1}{4\mathcal{T}}(\lambda\gamma_{m}\theta)(P^{m}+\mathcal{T}\partial_{\sigma}X^{m})+\tfrac{1}{16\mathcal{T}}(r\gamma^{m}\theta)(\lambda\gamma_{m}\theta)-\lambda^{\alpha}\partial_{\sigma}s_{\alpha}\nonumber \\
 &  & +A\lambda^{\alpha}s_{\alpha}-\Omega(\lambda^{\alpha}w_{\alpha}-r^{\alpha}s_{\alpha}+\theta^{\alpha}\pi_{\alpha}-\partial_{\sigma}B-\Sigma\chi-\beta\gamma)\nonumber \\
 &  & +\gamma(\lambda^{\alpha}p_{\alpha}-r^{\alpha}\epsilon_{\alpha}-\partial_{\sigma}\Sigma-A\Sigma)+\epsilon_{\alpha}\partial_{\sigma}\xi^{\alpha}-\chi\lambda^{\alpha}\epsilon_{\alpha},\\
\hat{J} & = & -\tfrac{1}{4\mathcal{T}}(\hat{\lambda}\gamma_{m}\hat{\theta})(P^{m}-\mathcal{T}\partial_{\sigma}X^{m})-\tfrac{1}{16\mathcal{T}}(\hat{r}\gamma^{m}\hat{\theta})(\hat{\lambda}\gamma_{m}\hat{\theta})-\hat{\lambda}^{\hat{\alpha}}\partial_{\sigma}\hat{s}_{\alpha}\nonumber \\
 &  & +\hat{A}\hat{\lambda}^{\hat{\alpha}}\hat{s}_{\alpha}-\hat{\Omega}(\hat{\lambda}^{\hat{\alpha}}\hat{w}_{\hat{\alpha}}+\hat{\theta}^{\hat{\alpha}}\hat{\pi}_{\hat{\alpha}}-\hat{r}^{\hat{\alpha}}\hat{s}_{\hat{\alpha}}-\partial_{\sigma}\hat{B}-\hat{\Sigma}\hat{\chi}-\hat{\beta}\hat{\gamma})\nonumber \\
 &  & +\hat{\gamma}(\hat{\lambda}^{\hat{\alpha}}\hat{p}_{\hat{\alpha}}-\hat{r}^{\hat{\alpha}}\hat{\epsilon}_{\hat{\alpha}}-\partial_{\sigma}\hat{\Sigma}-\hat{A}\hat{\Sigma})+\hat{\epsilon}_{\hat{\alpha}}\partial_{\sigma}\hat{\xi}^{\hat{\alpha}}-\hat{\chi}\hat{\lambda}^{\hat{\alpha}}\hat{\epsilon}_{\hat{\alpha}},
\end{eqnarray}
\end{subequations}such that $\partial_{-}J=\partial_{+}\hat{J}=0$.

As one final consistency check, observe that the generators of reparametrization
symmetry are exact. By defining,\begin{subequations}
\begin{eqnarray}
B_{+} & \equiv & \tfrac{(\Lambda\gamma^{m}\pi)}{(\Lambda\lambda)}[P_{m}+\mathcal{T}\partial_{\sigma}X_{m}-\tfrac{1}{2}(r\gamma_{m}\theta)]-r^{\alpha}w_{\alpha}+\eta^{\alpha}p_{\alpha}-\beta\chi-A\bar{\Omega},\\
B_{-} & \equiv & -\tfrac{(\hat{\Lambda}\gamma^{m}\hat{\pi})}{(\hat{\Lambda}\hat{\lambda})}[P_{m}-\mathcal{T}\partial_{\sigma}X_{m}+\tfrac{1}{2}(\hat{r}\gamma_{m}\hat{\theta})]-\hat{r}^{\hat{\alpha}}\hat{w}_{\hat{\alpha}}+\hat{\eta}^{\hat{\alpha}}\hat{p}_{\hat{\alpha}}-\hat{\beta}\hat{\chi}-\hat{A}\hat{\bar{\Omega}},
\end{eqnarray}
\end{subequations}it is straightforward to compute their BV-BRST
transformation, \emph{c.f.} equations \eqref{eq:BV-BRSTfields} and
\eqref{eq:BV-BRSTghosts},\begin{subequations}
\begin{eqnarray}
\delta B_{+} & = & -\tfrac{1}{4\mathcal{T}}(P_{m}+\mathcal{T}\partial_{\sigma}X_{m})(P^{m}+\mathcal{T}\partial_{\sigma}X^{m})-w_{\alpha}\partial_{\sigma}\lambda^{\alpha}-p_{\alpha}\partial_{\sigma}\xi^{\alpha}-A\partial_{\sigma}B\nonumber \\
 &  & -\chi\partial_{\sigma}\Sigma-\pi_{\alpha}\partial_{\sigma}\theta^{\alpha}-r^{\alpha}\partial_{\sigma}s_{\alpha}-\eta^{\alpha}\partial_{\sigma}\epsilon_{\alpha}-\beta\partial_{\sigma}\gamma-\bar{\Omega}\partial_{\sigma}\Omega,\\
\delta B_{-} & = & \tfrac{1}{4\mathcal{T}}(P^{m}-\mathcal{T}\partial_{\sigma}X^{m})(P_{m}-\mathcal{T}\partial_{\sigma}X_{m})-\hat{w}_{\hat{\alpha}}\partial_{\sigma}\hat{\lambda}^{\hat{\alpha}}-\hat{p}_{\hat{\alpha}}\partial_{\sigma}\hat{\xi}^{\hat{\alpha}}-\hat{A}\partial_{\sigma}\hat{B}\nonumber \\
 &  & -\hat{\chi}\partial_{\sigma}\hat{\Sigma}-\hat{\pi}_{\hat{\alpha}}\partial_{\sigma}\hat{\theta}^{\hat{\alpha}}-\hat{r}^{\hat{\alpha}}\partial_{\sigma}\hat{s}_{\hat{\alpha}}-\hat{\eta}^{\hat{\alpha}}\partial_{\sigma}\hat{\epsilon}_{\hat{\alpha}}-\hat{\beta}\partial_{\sigma}\hat{\gamma}-\hat{\bar{\Omega}}\partial_{\sigma}\hat{\Omega},
\end{eqnarray}
\end{subequations}which constitute the generalization of $H^{\pm}$
in \eqref{eq:H+-}.

\subsection{Field redefinition and emergent spacetime supersymmetry\label{subsec:Fieldred}}

Although not obviously, the BRST structure of the action \eqref{eq:PSfixed3}
can be greatly simplified. Consider the field redefinitions
\begin{equation}
\begin{array}{rclcrcl}
\lambda^{\alpha} & \to & \gamma^{-1}\lambda^{\alpha}, &  & p_{\alpha} & \to & p_{\alpha}+\partial_{\sigma}s_{\alpha},\\
w_{\alpha} & \to & \gamma w_{\alpha}, &  & \pi_{\alpha} & \to & \gamma^{-1}\pi_{\alpha},\\
A & \to & A+\frac{\partial_{\sigma}\gamma}{\gamma}, &  & r^{\alpha} & \to & \gamma^{-1}(r^{\alpha}+\partial_{\sigma}\xi^{\alpha}),\\
\chi & \to & \gamma\chi, &  & s_{\alpha} & \to & \gamma s_{\alpha},\\
\Sigma & \to & \gamma^{-1}\Sigma, &  & \beta & \to & \beta+\gamma^{-1}(\lambda^{\alpha}w_{\alpha}+\theta^{\alpha}\pi_{\alpha}-\partial_{\sigma}B)\\
\theta^{\alpha} & \to & \gamma\theta^{\alpha}, &  &  &  & +\gamma^{-1}(s_{\alpha}r^{\alpha}+s_{\alpha}\partial_{\sigma}\xi^{\alpha}+\chi\Sigma),
\end{array}\label{eq:fieldredefinition}
\end{equation}
and analogous operations in the hatted sector, which leave the action
\eqref{eq:PSfixed3} invariant. Note that the pure spinor constraints
\eqref{eq:PSconstraintsantighosts} have to be modified accordingly. 

Since the ghosts fields $\gamma$ and $\hat{\gamma}$ transform under
scaling, the field redefinitions above leave all the spacetime spinors
scale invariant at the price of shifting their ghost number. In particular,
the pure spinors $\lambda^{\alpha}$ and $\hat{\lambda}^{\hat{\alpha}}$
now have ghost number $+1$ while the $\theta^{\alpha}$ and $\hat{\theta}^{\hat{\alpha}}$
have ghost number zero. Also, due to the ordering of the operators,
the scaling parts of the BRST current (with $\Omega$ and $\hat{\Omega}$)
receive quantum corrections which can be cast as\begin{subequations}\label{eq:U(1)ordering}
\begin{eqnarray}
\Omega(\lambda^{\alpha}w_{\alpha}-r^{\alpha}s_{\alpha}+\theta^{\alpha}\pi_{\alpha}-\partial_{\sigma}B-\Sigma\chi-\beta\gamma) & \to & -\Omega(\beta\gamma+c_{\#}\partial_{\sigma}\ln\gamma),\\
\hat{\Omega}(\hat{\lambda}^{\hat{\alpha}}\hat{w}_{\hat{\alpha}}+\hat{\theta}^{\hat{\alpha}}\hat{\pi}_{\hat{\alpha}}-\hat{r}^{\hat{\alpha}}\hat{s}_{\hat{\alpha}}-\partial_{\sigma}\hat{B}-\hat{\Sigma}\hat{\chi}-\hat{\beta}\hat{\gamma}) & \to & -\hat{\Omega}(\hat{\beta}\hat{\gamma}+\hat{c}_{\#}\partial_{\sigma}\ln\hat{\gamma}),
\end{eqnarray}
\end{subequations}where $c_{\#}$ and $\hat{c}_{\#}$ are numerical
constants which will be fixed later in Appendix \ref{sec:decoupledsector}. 

The BRST currents \eqref{eq:PS-BRSTcurrent} are then rewritten as\begin{subequations}
\begin{eqnarray}
J & = & -\tfrac{1}{4\mathcal{T}}(\lambda\gamma_{m}\theta)\Big[P^{m}+\mathcal{T}\partial_{\sigma}X^{m}-\tfrac{1}{4}(\partial_{\sigma}\xi\gamma^{m}\theta)\Big]\nonumber \\
 &  & +\lambda^{\alpha}p_{\alpha}-\Omega(\beta\gamma+c_{\#}\partial_{\sigma}\ln\gamma)-A\Sigma-r^{\alpha}\epsilon_{\alpha}\nonumber \\
 &  & +\tfrac{1}{16\mathcal{T}}(r\gamma^{m}\theta)(\lambda\gamma_{m}\theta)+A\lambda^{\alpha}s_{\alpha}-\chi\lambda^{\alpha}\epsilon_{\alpha}-\partial_{\sigma}(\gamma\Sigma),\label{eq:JfieldR}\\
\hat{J} & = & -\tfrac{1}{4\mathcal{T}}(\hat{\lambda}\gamma_{m}\hat{\theta})\Big[P^{m}-\mathcal{T}\partial_{\sigma}X^{m}+\tfrac{1}{4}(\partial_{\sigma}\hat{\xi}\gamma^{m}\hat{\theta})\Big]\nonumber \\
 &  & +\hat{\lambda}^{\hat{\alpha}}\hat{p}_{\hat{\alpha}}-\hat{\Omega}(\hat{\beta}\hat{\gamma}+\hat{c}_{\#}\partial_{\sigma}\ln\hat{\gamma})-\hat{A}\hat{\Sigma}-\hat{r}^{\hat{\alpha}}\hat{\epsilon}_{\hat{\alpha}}\nonumber \\
 &  & -\tfrac{1}{16\mathcal{T}}(\hat{r}\gamma^{m}\hat{\theta})(\hat{\lambda}\gamma_{m}\hat{\theta})+\hat{A}\hat{\lambda}^{\hat{\alpha}}\hat{s}_{\alpha}-\hat{\chi}\hat{\lambda}^{\hat{\alpha}}\hat{\epsilon}_{\hat{\alpha}}-\partial_{\sigma}(\hat{\gamma}\hat{\Sigma}).\label{eq:JhatfieldR}
\end{eqnarray}
\end{subequations}The last term in each current can be disregarded,
since they correspond to total derivatives and do not contribute to
the BRST charges. Furthermore, the remaining terms in the third lines
of \eqref{eq:JfieldR} and \eqref{eq:JhatfieldR} can be removed by
similarity transformations of the form $J^{'}\equiv e^{-U}Je^{U}$
and $\hat{J}^{'}\equiv e^{-\hat{U}}\hat{J}e^{\hat{U}}$, where $U$
and $\hat{U}$ are invariant under the pure spinor symmetries and
given by\begin{subequations}
\begin{eqnarray}
U & = & \int d\sigma\Bigg\{\chi\lambda^{\alpha}s_{\alpha}-\tfrac{1}{32\mathcal{T}}(\lambda\gamma^{m}\theta)(\lambda\gamma^{n}\theta)\frac{(\eta\gamma_{mn}\Lambda)}{(\Lambda\lambda)}\Bigg\},\\
\hat{U} & = & \int d\sigma\Bigg\{\hat{\chi}\hat{\lambda}^{\hat{\alpha}}\hat{s}_{\hat{\alpha}}+\tfrac{1}{32\mathcal{T}}(\hat{\lambda}\gamma^{m}\hat{\theta})(\hat{\lambda}\gamma^{n}\hat{\theta})\frac{(\hat{\eta}\gamma_{mn}\hat{\Lambda})}{(\hat{\Lambda}\hat{\lambda})}\Bigg\},
\end{eqnarray}
\end{subequations}It is then straightforward to show that\begin{subequations}
\begin{eqnarray}
J^{'} & = & \lambda^{\alpha}p_{\alpha}-\tfrac{1}{4\mathcal{T}}(\lambda\gamma_{m}\theta)\Big[P^{m}+\mathcal{T}\partial_{\sigma}X^{m}-\tfrac{1}{4}(\partial_{\sigma}\xi\gamma^{m}\theta)\Big]\nonumber \\
 &  & -\Omega(\beta\gamma+c_{\#}\partial_{\sigma}\ln\gamma)-A\Sigma-r^{\alpha}\epsilon_{\alpha},\\
\hat{J}^{'} & = & \hat{\lambda}^{\hat{\alpha}}\hat{p}_{\hat{\alpha}}-\tfrac{1}{4\mathcal{T}}(\hat{\lambda}\gamma_{m}\hat{\theta})\Big[P^{m}-\mathcal{T}\partial_{\sigma}X^{m}+\tfrac{1}{4}(\partial_{\sigma}\hat{\xi}\gamma^{m}\hat{\theta})\Big]\nonumber \\
 &  & -\hat{\Omega}(\hat{\beta}\hat{\gamma}+\hat{c}_{\#}\partial_{\sigma}\ln\hat{\gamma})-\hat{A}\hat{\Sigma}-\hat{r}^{\hat{\alpha}}\hat{\epsilon}_{\hat{\alpha}}.
\end{eqnarray}
\end{subequations}

It is important to emphasize that the field redefinitions in \eqref{eq:fieldredefinition}
are well defined only if $\gamma$ and $\hat{\gamma}$ are assumed
to be  non-vanishing in every point of the worldsheet. This is clear
in the definition of the conformal field theory of the decoupled sector,
which is singular for $\gamma=0$ and $\hat{\gamma}=0$. In a path
integral formulation, this singularity can be avoided by choosing
a convenient parametrization for the ghosts, \emph{e.g.} $\gamma=e^{\sigma}$
and $\beta=\rho e^{-\sigma}$, therefore enforcing the non-vanishing
condition. Here, $\sigma$ is a chiral worldsheet scalar with conjugate
$\rho$. More details can be found in the appendix \ref{sec:decoupledsector}. 

Using the quartet argument, the BRST cohomology can be shown to be
independent of $A$, $B$, $\chi$, $\sigma$, $\rho$, $\Sigma$,
$r^{\alpha}$, $s_{\alpha}$, $\eta^{\alpha}$ and $\epsilon_{\alpha}$
(hatted and unhatted). Therefore, these fields can be eliminated from
the theory and the gauge fixed action \eqref{eq:PSfixed3} is further
simplified to
\begin{eqnarray}
\tilde{S} & = & \int d\tau d\sigma\{P_{m}\partial_{\tau}X^{m}-\tfrac{1}{2\mathcal{T}}(P_{m}P^{m}+\mathcal{T}^{2}\partial_{\sigma}X_{m}\partial_{\sigma}X^{m})\}\nonumber \\
 &  & +\int d\tau d\sigma\{w_{\alpha}\partial_{-}\lambda^{\alpha}+p_{\alpha}\partial_{-}\xi^{\alpha}+\pi_{\alpha}\partial_{-}\theta^{\alpha}+\hat{w}_{\hat{\alpha}}\partial_{+}\hat{\lambda}^{\hat{\alpha}}+\hat{p}_{\hat{\alpha}}\partial_{+}\hat{\xi}^{\hat{\alpha}}+\hat{\pi}_{\hat{\alpha}}\partial_{+}\hat{\theta}^{\hat{\alpha}}\}.\label{eq:PSfixed}
\end{eqnarray}

The pure spinor constraints \eqref{eq:PSconstraintsBV} are reduced
to
\begin{equation}
\begin{array}{rclcrcl}
(\lambda\gamma^{m}\lambda) & = & 0, &  & (\hat{\lambda}\gamma^{m}\hat{\lambda}) & = & 0,\\
(\lambda\gamma^{m}\xi) & = & 0, &  & (\hat{\lambda}\gamma^{m}\hat{\xi}) & = & 0,\\
\xi^{\alpha}\pi_{\alpha} & = & 0, &  & \hat{\xi}^{\hat{\alpha}}\hat{\pi}_{\hat{\alpha}} & = & 0,\\
\lambda^{\alpha}\pi_{\alpha} & = & 0, &  & \hat{\lambda}^{\hat{\alpha}}\hat{\pi}_{\hat{\alpha}} & = & 0,\\
(\lambda\gamma^{mn}\pi) & = & 0, &  & (\hat{\lambda}\gamma^{mn}\hat{\pi}) & = & 0,
\end{array}\label{eq:PSconstraintsBRST}
\end{equation}
and the action \eqref{eq:PSfixed} is invariant under the implied
pure spinor gauge transformations\begin{subequations}\label{eq:PSsymmetriesBRST}
\begin{eqnarray}
\delta w_{\alpha} & = & d_{m}(\gamma^{m}\lambda)_{\alpha}+e_{m}(\gamma^{m}\xi)_{\alpha}-\bar{f}\pi_{\alpha}-\bar{f}_{mn}(\gamma^{mn}\pi)_{\alpha},\\
\delta p_{\alpha} & = & e_{m}(\gamma^{m}\lambda)_{\alpha}-\bar{g}\pi_{\alpha},\\
\delta\theta^{\alpha} & = & \bar{f}\lambda^{\alpha}+\bar{f}_{mn}(\gamma^{mn}\lambda)^{\alpha}+\bar{g}\xi^{\alpha},\\
\delta\hat{w}_{\hat{\alpha}} & = & \hat{d}_{m}(\gamma^{m}\hat{\lambda})_{\hat{\alpha}}+\hat{e}_{m}(\gamma^{m}\hat{\xi})_{\hat{\alpha}}-\hat{\bar{f}}\hat{\pi}_{\hat{\alpha}}-\hat{\bar{f}}_{mn}(\gamma^{mn}\hat{\pi})_{\hat{\alpha}},\\
\delta\hat{p}_{\hat{\alpha}} & = & \hat{e}_{m}(\gamma^{m}\hat{\lambda})_{\hat{\alpha}}-\hat{\bar{g}}\hat{\pi}_{\hat{\alpha}},\\
\delta\hat{\theta}^{\hat{\alpha}} & = & \hat{\bar{f}}\hat{\lambda}^{\hat{\alpha}}+\hat{\bar{f}}_{mn}(\gamma^{mn}\hat{\lambda})^{\hat{\alpha}}+\hat{\bar{g}}\hat{\xi}^{\hat{\alpha}}.
\end{eqnarray}
\end{subequations}

These transformations are the key to spacetime supersymmetry. The
parameter $e_{m}$ can be tuned in such a way that the gauge dependent
components of $p_{\alpha}$ are identified with the independent components
of the constrained spinor $\pi_{\alpha}$. Furthermore, the gauge
parameters $\bar{f}$ and $\bar{f}_{mn}$ can be similarly chosen
such that the gauge dependent components of $\theta^{\alpha}$ are
identified with the independent components of the constrained spinor
$\xi^{\alpha}$. This is demonstrated in Appendix \ref{sec:Partial-gauge-fixing}.
An analogous gauge fixing can be performed in the hatted sector. The
outcome of the partial gauge fixing of the pure spinor symmetries
is the action
\begin{eqnarray}
S & = & \int d\tau d\sigma\{P_{m}\partial_{\tau}X^{m}-\tfrac{1}{2\mathcal{T}}(P_{m}P^{m}+\mathcal{T}^{2}\partial_{\sigma}X_{m}\partial_{\sigma}X^{m})\}\nonumber \\
 &  & +\int d\tau d\sigma\{w_{\alpha}\partial_{-}\lambda^{\alpha}+p_{\alpha}\partial_{-}\theta^{\alpha}+\hat{w}_{\hat{\alpha}}\partial_{+}\hat{\lambda}^{\hat{\alpha}}+\hat{p}_{\hat{\alpha}}\partial_{+}\hat{\theta}^{\hat{\alpha}}\},\label{eq:PSaction}
\end{eqnarray}
but now with unconstrained $p_{\alpha}$, $\theta^{\alpha}$, $\hat{p}_{\hat{\alpha}}$
and $\hat{\theta}^{\hat{\alpha}}$. The action $S$ corresponds to
the type II-B superstring in the pure spinor formalism. The type II-A
is similarly obtained by reverting the spinor chirality of one of
the sectors, either hatted or unhatted. The heterotic superstring
is obtained when only one of the twistor-like constraints \eqref{eq:twistorconstraints}
is imposed, but then worldsheet reparametrization has be taken into
account in the construction of the master action.

The non-vanishing BRST transformations can be cast as
\begin{equation}
\begin{array}{rclcrcl}
\delta X^{m} & = & \tfrac{1}{4\mathcal{T}}(\lambda\gamma^{m}\theta)+\tfrac{1}{4\mathcal{T}}(\hat{\lambda}\gamma^{m}\hat{\theta}), &  & \delta P_{m} & = & \tfrac{1}{4}\partial_{\sigma}[(\lambda\gamma_{m}\theta)]-\tfrac{1}{4}\partial_{\sigma}[(\hat{\lambda}\gamma_{m}\hat{\theta})],\\
\delta w_{\alpha} & = & p_{\alpha}-\tfrac{1}{16\mathcal{T}}(\theta\gamma_{m}\partial_{\sigma}\theta)(\gamma^{m}\theta)_{\alpha} &  & \delta\hat{w}_{\hat{\alpha}} & = & \hat{p}_{\hat{\alpha}}+\tfrac{1}{16\mathcal{T}}(\hat{\theta}\gamma_{m}\partial_{\sigma}\hat{\theta})(\gamma^{m}\hat{\theta})_{\hat{\alpha}}\\
 &  & -\tfrac{1}{4\mathcal{T}}(P_{m}+\mathcal{T}\partial_{\sigma}X_{m})(\gamma^{m}\theta)_{\alpha}, &  &  &  & -\tfrac{1}{4\mathcal{T}}(P_{m}-\mathcal{T}\partial_{\sigma}X_{m})(\gamma^{m}\hat{\theta})_{\hat{\alpha}},\\
\delta\theta^{\alpha} & = & \lambda^{\alpha}, &  & \delta\hat{\theta}^{\hat{\alpha}} & = & \hat{\lambda}^{\hat{\alpha}},\\
\delta p_{\alpha} & = & -\tfrac{1}{4\mathcal{T}}(P_{m}+\mathcal{T}\partial_{\sigma}X_{m})(\gamma^{m}\lambda)_{\alpha} &  & \delta\hat{p}_{\hat{\alpha}} & = & -\tfrac{1}{4\mathcal{T}}(P_{m}-\mathcal{T}\partial_{\sigma}X_{m})(\gamma^{m}\hat{\lambda})_{\hat{\alpha}}\\
 &  & +\tfrac{1}{16\mathcal{T}}(\partial_{\sigma}\lambda\gamma_{m}\theta)(\gamma^{m}\theta)_{\alpha} &  &  &  & -\tfrac{1}{16\mathcal{T}}(\partial_{\sigma}\hat{\lambda}\gamma_{m}\hat{\theta})(\gamma^{m}\hat{\theta})_{\hat{\alpha}}\\
 &  & +\tfrac{3}{8\mathcal{T}}(\lambda\gamma_{m}\theta)(\gamma^{m}\partial_{\sigma}\theta)_{\alpha} &  &  &  & -\tfrac{3}{8\mathcal{T}}(\hat{\lambda}\gamma_{m}\hat{\theta})(\gamma^{m}\partial_{\sigma}\hat{\theta})_{\hat{\alpha}}
\end{array}
\end{equation}
generated by the BRST charges\begin{subequations}
\begin{eqnarray}
Q & = & \int d\sigma\{\lambda^{\alpha}p_{\alpha}-\tfrac{1}{4\mathcal{T}}(\lambda\gamma^{m}\theta)(P_{m}+\mathcal{T}\partial_{\sigma}X_{m})+\tfrac{1}{16\mathcal{T}}(\lambda\gamma^{m}\theta)(\theta\gamma_{m}\partial_{\sigma}\theta)\},\label{eq:PSBRSTholo}\\
\hat{Q} & = & \int d\sigma\{\hat{\lambda}^{\hat{\alpha}}\hat{p}_{\hat{\alpha}}-\tfrac{1}{4\mathcal{T}}(\hat{\lambda}\gamma^{m}\hat{\theta})(P_{m}-\mathcal{T}\partial_{\sigma}X_{m})-\tfrac{1}{16\mathcal{T}}(\hat{\lambda}\gamma^{m}\hat{\theta})(\hat{\theta}\gamma_{m}\partial_{\sigma}\hat{\theta})\},\label{eq:PSBRSTantiholo}
\end{eqnarray}
\end{subequations}

Finally, note that the BRST charges and the action \eqref{eq:PSaction}
are spacetime supersymmetric and the supersymmetry generators can
be expressed as\begin{subequations}\label{eq:SUSYcharges}
\begin{eqnarray}
q_{\alpha} & = & \int d\sigma\Big\{ p_{\alpha}+\tfrac{1}{4\mathcal{T}}(P_{m}+\mathcal{T}\partial_{\sigma}X_{m})(\gamma^{m}\theta)_{\alpha}+\tfrac{1}{48\mathcal{T}}(\theta\gamma_{m}\partial_{\sigma}\theta)(\gamma^{m}\theta)_{\alpha}\Big\},\\
\hat{q}_{\hat{\alpha}} & = & \int d\sigma\Big\{\hat{p}_{\hat{\alpha}}+\tfrac{1}{4\mathcal{T}}(P_{m}-\mathcal{T}\partial_{\sigma}X_{m})(\gamma^{m}\hat{\theta})_{\hat{\alpha}}-\tfrac{1}{48\mathcal{T}}(\hat{\theta}\gamma_{m}\partial_{\sigma}\hat{\theta})(\gamma^{m}\hat{\theta})_{\hat{\alpha}}\Big\},
\end{eqnarray}
\end{subequations}satisfying the algebra\begin{subequations}\label{eq:SUSYalgebra}
\begin{eqnarray}
\{q_{\alpha},q_{\beta}\} & = & \tfrac{1}{2\mathcal{T}}\gamma_{\alpha\beta}^{m}\int d\sigma\,P_{m},\\
\{\hat{q}_{\hat{\alpha}},\hat{q}_{\hat{\beta}}\} & = & \tfrac{1}{2\mathcal{T}}\gamma_{\hat{\alpha}\hat{\beta}}^{m}\int d\sigma\,P_{m}.
\end{eqnarray}
\end{subequations}

Therefore, spacetime supersymmetry is made manifest with the help
of the field redefinitions \eqref{eq:fieldredefinition} and the pure
spinor gauge symmetries \eqref{eq:PSsymmetriesBRST}, demonstrating
 the the action \eqref{eq:newaction} can be seen as the underlying
gauge theory of the pure spinor superstring

\section{Final remarks\label{sec:Final-remarks}}

The pure spinor action \eqref{eq:PSaction} can be rewritten in a
more traditional way by solving the equation of motion for $P_{m}$
and Wick-rotating the worldsheet time coordinate $\tau$. The resulting
action is
\begin{equation}
S=\int d^{2}z\{\tfrac{1}{2}\partial X^{m}\bar{\partial}X_{m}+w_{\alpha}\bar{\partial}\lambda^{\alpha}+p_{\alpha}\bar{\partial}\theta^{\alpha}+\hat{w}_{\hat{\alpha}}\partial\hat{\lambda}^{\hat{\alpha}}+\hat{p}_{\hat{\alpha}}\partial\hat{\theta}^{\hat{\alpha}}\},\label{eq:PSactiontraditional}
\end{equation}
with $z$ ($\bar{z}$) denoting the usual (anti)holomorphic coordinate,
$\partial\equiv\tfrac{\partial}{\partial z}$, $\bar{\partial}\equiv\tfrac{\partial}{\partial\bar{z}}$
and $\mathcal{T}=1$ (string tension). The BRST charge \eqref{eq:PSBRSTholo}
takes its standard form in the pure spinor formalism as
\begin{equation}
Q_{\tiny{PS}}=\ointctrclockwise\,\lambda^{\alpha}d_{\alpha},\label{eq:PSBRSTcharge}
\end{equation}
where $d_{\alpha}$ denotes the field realization of the supersymmetric
derivative and is expressed as
\begin{equation}
d_{\alpha}\equiv p_{\alpha}-\tfrac{1}{2}\partial X^{m}(\gamma_{m}\theta)_{\alpha}-\tfrac{1}{8}(\theta\gamma^{m}\partial\theta)(\gamma_{m}\theta)_{\alpha}.
\end{equation}

While the twistor-like symmetry can be understood as a way to rewrite
the generators of worldsheet reparametrization with a linear dependence
on $P_{m}$, the extra fermionic pair $\{\xi^{\alpha},p_{\alpha}\}$
and  the fermionic symmetry \eqref{eq:fermionSYM} do not have a clear
physical interpretation. These ingredients are ultimately responsible
for the emergence of spacetime supersymmetry but from the worldsheet
point of view their existence lacks a more fundamental understanding.
The operator $\lambda^{\alpha}p_{\alpha}$ resembles part of a possible
worldsheet supersymmetry generator with ``wrong'' conformal dimension
so it might be possible that the gauge action \eqref{eq:newaction}
can be embedded in a bigger model with twisted worldsheet supersymmetry.
It would be interesting to investigate potential connections with
(1) the superembedding origin of the heterotic pure spinor superstring,
presented in \cite{Matone:2002ft}; and (2) the twisted formulation
of the pure spinor superstring introduced in \cite{Berkovits:2016xnb}
and its relation to the spinning string. Also in this direction, it
is worth studying alternative gauge choices for the master action
\eqref{eq:PSMASTER} and whether they could be related to the Green-Schwarz
superstring. This idea was first proposed in \cite{Berkovits:2015yra}
for Berkovits' action \eqref{eq:Berkovitsaction} and it would be
interesting to develop  a similar approach here. In order to do that,
it seems that worldsheet reparametrization has to be explicitly included
in the construction of the master action.

\textbf{Acknowledgments:} I would like to thank Thales Azevedo and
Nathan Berkovits for comments of the draft. Also, I would like to
thank Dmitri Sorokin for reference suggestions and for pointing out
a possible connection between the results presented here and the superembedding
approach. This research has been supported by the Czech Science Foundation
- GA\v{C}R, project 19-06342Y.

\appendix

\section{Partial gauge fixing of the pure spinor symmetries\label{sec:Partial-gauge-fixing}}

The aim of this appendix is to demonstrate that the action \eqref{eq:PSfixed}
can be rewritten in terms of unconstrained spacetime spinors $p_{\alpha}$,
$\theta^{\alpha}$, $\hat{p}_{\hat{\alpha}}$ and $\hat{\theta}^{\hat{\alpha}}$,
provided that the pure spinor gauge symmetries \eqref{eq:PSsymmetriesBRST}
are partially fixed in a precise form. In order to do this, the pure
spinor constraints \eqref{eq:PSconstraintsBRST} will be explicitly
solved in a Wick-rotated scenario and the $SO(10)$ spinors will be
expressed in terms of $U(5)$ components. To illustrate the procedure,
only the unhatted (left-moving) sector will be analyzed, but it can
be easily extended to the right-moving sector.

\subsection*{$U(5)$ decomposition}

Given an $SO\left(10\right)$ chiral spinor $\xi^{\alpha}$ (antichiral
$\chi_{\alpha}$), with $\alpha=1,\ldots,16$, it is possible to determine
its $U\left(5\right)$ components using the projectors $P_{A}^{\alpha}$
and $\left(P_{A}^{\alpha}\right)^{-1}\equiv P_{\alpha}^{I}$, where
$A=\{+,a,ab\}$ denotes the $U\left(5\right)$ indices, respectively
the $U(1)$-charged singlet, the vector and the adjoint representations,
with $a=1,\ldots,5$, defined in such a way that
\begin{equation}
\begin{array}{rclcrcl}
\xi^{\alpha} & = & P_{+}^{\alpha}\xi^{+}+\frac{1}{2}P_{ab}^{\alpha}\xi^{ab}+P^{\alpha a}\xi_{a}, &  & \chi_{\alpha} & = & P_{\alpha}^{+}\chi_{+}+\frac{1}{2}P_{\alpha}^{ab}\chi_{ab}+P_{\alpha a}\chi^{a},\\
\xi^{+} & = & P_{\alpha}^{+}\xi^{\alpha}, &  & \chi_{+} & = & P_{+}^{\alpha}\chi_{\alpha},\\
\xi^{ab} & = & P_{\alpha}^{ab}\xi^{\alpha}, &  & \chi_{ab} & = & P_{ab}^{\alpha}\chi_{\alpha},\\
\xi_{a} & = & P_{\alpha a}\xi^{\alpha}, &  & \chi^{a} & = & P^{\alpha a}\chi_{\alpha}.
\end{array}\label{eq:u5spinordecomposition}
\end{equation}
These projectors satisfy the orthogonality equations
\begin{equation}
\begin{array}{rcl}
P_{+}^{\alpha}P_{\alpha}^{+} & = & 1,\\
P^{\alpha a}P_{\alpha b} & = & \delta_{b}^{a},\\
P_{ab}^{\alpha}P_{\alpha}^{cd} & = & \delta_{a}^{c}\delta_{b}^{d}-\delta_{b}^{c}\delta_{a}^{d}\\
P_{+}^{\alpha}P_{\beta}^{+}+\frac{1}{2}P_{ab}^{\alpha}P_{\beta}^{ab}+P^{\alpha a}P_{\beta a} & = & \delta_{\beta}^{\alpha},
\end{array}\label{eq:orthoprojectorsSO(10)}
\end{equation}
and more generally $P_{A}^{\alpha}P_{\alpha}^{B}=0$ for $A\neq B$.
They can be used as building blocks of the symmetric $g$-matrices,
defined as
\begin{equation}
\begin{array}{rcl}
(g^{a})^{\alpha\beta} & \equiv & \sqrt{2}(P_{+}^{\alpha}P^{\beta a}+P_{+}^{\beta}P^{\alpha a}+\frac{1}{4}\epsilon^{abcde}P_{bc}^{\alpha}P_{de}^{\beta}),\\
(\bar{g}_{a})^{\alpha\beta} & \equiv & \sqrt{2}(P^{\alpha b}P_{ba}^{\beta}+P^{\beta b}P_{ba}^{\alpha}),\\
(g^{a})_{\alpha\beta} & \equiv & \sqrt{2}(P_{\beta b}P_{\alpha}^{ba}+P_{\alpha b}P_{\beta}^{ba}),\\
(\bar{g}_{a})_{\alpha\beta} & \equiv & \sqrt{2}(P_{\alpha}^{+}P_{\beta a}+P_{\beta}^{+}P_{\alpha a}+\frac{1}{4}\epsilon_{abcde}P_{\alpha}^{bc}P_{\beta}^{de}),
\end{array}\label{eq:gmatricesSO(10)}
\end{equation}
where $\epsilon^{abcde}$ and $\epsilon_{abcde}$ are the totally
antisymmetric $U\left(5\right)$ tensors, with $\epsilon^{12345}=\epsilon_{54321}=1$,
satisfying the algebra\begin{subequations}
\begin{eqnarray}
\{g^{a},g^{b}\}_{\alpha}^{\phantom{\alpha}\beta} & \equiv & g_{\alpha\gamma}^{a}(g^{b})^{\gamma\beta}+g_{\alpha\gamma}^{b}(g^{a})^{\gamma\beta}\nonumber \\
 & = & 0,\\
\{\bar{g}_{a},\bar{g}_{b}\}_{\alpha}^{\phantom{\alpha}\beta} & \equiv & (\bar{g}_{a})_{\alpha\gamma}\bar{g}_{b}^{\gamma\beta}+(\bar{g}_{b})_{\alpha\gamma}\bar{g}_{a}^{\gamma\beta}\nonumber \\
 & = & 0,\\
\{g^{a},\bar{g}_{b}\}_{\alpha}^{\phantom{\alpha}\beta} & \equiv & g_{\alpha\gamma}^{a}\bar{g}_{b}^{\gamma\beta}+(\bar{g}_{b})_{\alpha\gamma}(g^{a})^{\gamma\beta}\nonumber \\
 & = & 2\delta^{\beta}\delta_{b}^{a}.
\end{eqnarray}
\end{subequations}Therefore, the $g$-matrices \eqref{eq:gmatricesSO(10)}
represent the $U(5)$ components of the gamma matrices $\gamma^{m}$.

\subsection*{Solving the pure spinor constraints}

The action \eqref{eq:PSfixed} is invariant under the pure spinor
gauge transformations
\begin{equation}
\begin{array}{rcl}
\delta w_{\alpha} & = & d_{m}(\gamma^{m}\lambda)_{\alpha}+e_{m}(\gamma^{m}\xi)_{\alpha}-\bar{f}_{mn}(\gamma^{mn}\pi)_{\alpha},\\
\delta p_{\alpha} & = & e_{m}(\gamma^{m}\lambda)_{\alpha},\\
\delta\theta^{\alpha} & = & \bar{f}_{mn}(\gamma^{mn}\lambda)^{\alpha},
\end{array}\label{eq:PSgaugeAP}
\end{equation}
where $d_{m},$ $e_{m}$ and $\bar{f}_{mn}$ are the parameters. Note
that the parameters $\bar{f}$ and $\bar{g}$ in \eqref{eq:PSsymmetriesBRST}
are redundant, for they can be absorbed by a shift of the other parameters:
\begin{equation}
\begin{array}{rcl}
d_{m} & \to & d_{m}-\tfrac{4\bar{f}}{5}\frac{(\Lambda\gamma_{m}\pi)}{(\Lambda\lambda)}+\tfrac{\bar{g}}{5}\frac{(\Lambda\xi)}{(\Lambda\lambda)^{2}}(\Lambda\gamma_{m}\pi),\\
e_{m} & \to & e_{m}+\tfrac{\bar{g}}{2}\frac{(\Lambda\gamma_{m}\pi)}{(\Lambda\lambda)},\\
\bar{f}_{mn} & \to & \bar{f}_{mn}+\tfrac{\bar{f}}{10}\frac{(\Lambda\gamma_{mn}\lambda)}{(\Lambda\lambda)}+\tfrac{\bar{g}}{8}\frac{(\Lambda\gamma_{mn}\xi)}{(\Lambda\lambda)}-\tfrac{\bar{g}}{40}\frac{(\Lambda\xi)(\Lambda\gamma_{mn}\lambda)}{(\Lambda\lambda)^{2}}.
\end{array}
\end{equation}
This can be easily demonstrated with the help of the gamma matrix
property \eqref{eq:gammaproperty}.

In terms of $U(5)$ components, $\lambda^{\alpha}$, $\theta^{\alpha}$
and $\pi_{\alpha}$ can be parametrized as $\lambda^{\alpha}=(\lambda^{+},\lambda^{ab},\lambda_{a})$,
$\xi^{\alpha}=(\xi^{+},\xi^{ab},\xi_{a})$, and $\pi_{\alpha}=(\pi_{+},\pi_{ab},\pi^{a})$.
Therefore, using the $g$-matrices \eqref{eq:gmatricesSO(10)}, the
constraints \eqref{eq:PSconstraintsBRST} can be rewritten as
\begin{equation}
\begin{array}{rclcrcl}
\lambda^{ab}\lambda_{b} & = & 0, &  & \lambda^{+}\pi_{ab} & = & \frac{1}{2}\epsilon_{abcde}\pi^{c}\lambda^{de},\\
\lambda^{+}\lambda_{a} & = & -\tfrac{1}{8}\epsilon_{abcde}\lambda^{bc}\lambda^{de}, &  & \lambda^{+}\pi_{+}- & = & \tfrac{3}{5}\lambda_{a}\pi^{a}-\tfrac{1}{10}\lambda^{ab}\pi_{ab},\\
\lambda^{ab}\xi_{b} & = & -\xi^{ab}\lambda_{b}, &  & \lambda_{a}\pi^{b}-\lambda^{bc}\pi_{ac} & = & \tfrac{1}{5}\delta_{a}^{b}(\lambda_{c}\pi^{c}-\lambda^{cd}\pi_{cd}),\\
\lambda^{+}\xi_{a} & = & -\xi^{+}\lambda_{a}-\tfrac{1}{4}\epsilon_{abcde}\lambda^{bc}\xi^{de}, &  & \lambda^{+}\pi_{+}- & = & \lambda_{a}\pi^{a}+\tfrac{1}{2}\lambda^{ab}\pi_{ab},\\
\lambda^{ab}\pi_{+} & = & \frac{1}{2}\epsilon^{abcde}\lambda_{c}\pi_{de}, &  & \xi^{+}\pi_{+}- & = & \xi_{a}\pi^{a}+\tfrac{1}{2}\xi^{ab}\pi_{ab}.
\end{array}
\end{equation}

Assuming $\lambda^{+}\neq0$, these constraints can be explicitly
solved by
\begin{equation}
\begin{array}{rcl}
\lambda_{a} & = & -\tfrac{1}{8\lambda^{+}}\epsilon_{abcde}\lambda^{bc}\lambda^{de},\\
\xi_{a} & = & -\tfrac{\xi^{+}}{\lambda^{+}}\lambda_{a}-\tfrac{1}{4\lambda^{+}}\epsilon_{abcde}\lambda^{bc}\xi^{de},\\
\pi_{+} & = & \tfrac{1}{\lambda^{+}}\lambda_{a}\pi^{a},\\
\pi_{ab} & = & \tfrac{1}{2\lambda^{+}}\epsilon_{abcde}\pi^{c}\lambda^{de},
\end{array}\label{eq:PSsolutions}
\end{equation}

Next, the transformations \eqref{eq:PSgaugeAP} can be used to conveniently
choose the gauge $p_{\alpha}=\{p_{+},p_{ab},\pi^{a}\}$ and $\theta^{\alpha}=\{\xi^{+},\xi^{ab},\theta_{a}\}$,
where $p_{+}$, $p_{ab}$ and $\theta_{a}$ denote the independent
components of $p_{\alpha}$ and $\theta^{\alpha}$. Using this gauge
and the solutions \eqref{eq:PSsolutions}, their contribution to the
action \eqref{eq:PSfixed} can be cast as
\begin{eqnarray}
S_{sp} & = & \int d\tau d\sigma\{p_{\alpha}\partial_{-}\xi^{\alpha}+\pi_{\alpha}\partial_{-}\theta^{\alpha}\},\nonumber \\
 & = & \int d\tau d\sigma\{p_{+}\partial_{-}\xi^{+}+\tfrac{1}{2}p_{ab}\partial_{-}\xi^{ab}+p^{a}\partial_{-}\xi_{a}\}\nonumber \\
 &  & +\int d\tau d\sigma\{\pi_{+}\partial_{-}\theta^{+}+\tfrac{1}{2}\pi_{ab}\partial_{-}\theta^{ab}+\pi^{a}\partial_{-}\theta_{a}\},\nonumber \\
 & = & \int d\tau d\sigma\Big\{ p_{\alpha}\partial_{-}\theta^{\alpha}-\pi^{a}\xi^{+}\partial_{-}\Big(\tfrac{\lambda_{a}}{\lambda^{+}}\Big)-\tfrac{1}{4}\epsilon_{abcde}\pi^{a}\xi^{de}\partial_{-}\Big(\tfrac{\lambda^{bc}}{\lambda^{+}}\Big)\Big\}.
\end{eqnarray}
Note that the last two terms can be absorbed by a redefinition of
$w_{\alpha}$ and the resulting action \eqref{eq:PSaction} depends
only on the unconstrained pair $\{p_{\alpha},\theta^{\alpha}\}$.

Furthermore, the BRST current derived from the action \eqref{eq:PSBRSTholo}
is rewritten as well in terms of  the pair $\{p_{\alpha},\theta^{\alpha}\}$.
The terms $\lambda^{\alpha}p_{\alpha}$ and $(\lambda\gamma^{m}\theta)$
preserve their shape with the gauge choice above and after a simple
similarity transformation, the BRST current is given by
\begin{equation}
J=\lambda^{\alpha}p_{\alpha}-\tfrac{1}{4\mathcal{T}}(\lambda\gamma_{m}\theta)\Big[P^{m}+\mathcal{T}\partial_{\sigma}X^{m}+\tfrac{1}{4}(\theta\gamma^{m}\partial_{\sigma}\theta)\Big],
\end{equation}
corresponding to the usual pure spinor BRST current plus a $U(1)$
decoupled sector.

\section{The $U(1)_{R}\times U(1)_{L}$ sector\label{sec:decoupledsector}}

This section presents some properties of the $U(1)_{R}\times U(1)_{L}$
sector, which is constituted by the ghosts coming from the scaling
symmetry \eqref{eq:scalingSYM} and the fermionic symmetry \eqref{eq:fermionSYM},
but after the field redefinition \eqref{eq:fieldredefinition}. 

After a Wick-rotation of the worldsheet time $\tau$, their contribution
to the action \eqref{eq:PSfixed3} can be written as
\begin{equation}
S_{*}=\int d^{2}z\{\bar{\Omega}\bar{\partial}\Omega+\beta\bar{\partial}\gamma+\hat{\bar{\Omega}}\partial\hat{\Omega}+\hat{\beta}\partial\hat{\gamma}\},\label{eq:BU1action}
\end{equation}
with holomorphic and anti-holomorphic BRST currents given by\begin{subequations}\label{eq:BU1BRST}
\begin{eqnarray}
J_{*} & = & \Omega\big(\tfrac{1}{2}\partial\ln\gamma-\beta\gamma\big),\\
\hat{J}_{*} & = & \hat{\Omega}\big(\tfrac{1}{2}\bar{\partial}\ln\hat{\gamma}-\hat{\beta}\hat{\gamma}\big).
\end{eqnarray}
\end{subequations}Note that the constants $c_{\#}$ and $\hat{c}_{\#}$
in \eqref{eq:U(1)ordering} were fixed by requiring nilpotency of
the BRST charges $Q_{*}\equiv\ointctrclockwise\,J_{*}$ and $\hat{Q}_{*}\equiv\varointclockwise\,\hat{J}_{*}$.

The cohomology of $Q_{*}$ has only two elements, the identity operator
$\mathbbm{1}$ and $\Omega$, which is BRST singular,
\begin{equation}
\Omega=\lim_{\epsilon\to0}\tfrac{1}{\epsilon}[Q_{*},\gamma^{\epsilon}].
\end{equation}
Any other BRST-closed operator can be shown to be $Q_{*}$-exact and
there is no operator trivializing the cohomology.

Because of the field redefinition \eqref{eq:fieldredefinition}, $\beta$
and $\hat{\beta}$ are not conformal primary operators. In fact, the
components of the energy-momentum tensor have a non-standard form
given by\begin{subequations}
\begin{eqnarray}
T_{*} & = & -\bar{\Omega}\partial\Omega-\beta\partial\gamma+\tfrac{1}{2\gamma^{2}}[\gamma\partial^{2}\gamma-(\partial\gamma)^{2}],\\
\hat{T}_{*} & = & -\hat{\bar{\Omega}}\bar{\partial}\hat{\Omega}-\hat{\beta}\bar{\partial}\hat{\gamma}+\tfrac{1}{2\hat{\gamma}^{2}}[\hat{\gamma}\bar{\partial}^{2}\hat{\gamma}-(\bar{\partial}\hat{\gamma})^{2}].
\end{eqnarray}
\end{subequations}

Using the fundamental OPE
\begin{equation}
\gamma(z)\,\beta(y)\sim\frac{1}{(z-y)},
\end{equation}
the following results are obtained\begin{subequations}
\begin{eqnarray}
T_{*}(z)\,\gamma(y) & \sim & \frac{\partial\gamma}{(z-y)},\\
T_{*}(z)\,\beta(y) & \sim & \frac{\gamma^{-1}}{(z-y)^{3}}+\frac{\beta}{(z-y)^{2}}+\frac{\partial\beta}{(z-y)},\\
T_{*}(z)\,\Omega(y) & \sim & \frac{\partial\Omega}{(z-y)},\\
T_{*}(z)\,\bar{\Omega}(y) & \sim & \frac{\bar{\Omega}}{(z-y)^{2}}+\frac{\partial\bar{\Omega}}{(z-y)},\\
T_{*}(z)\,J_{*}(y) & \sim & \frac{J_{*}}{(z-y)^{2}}+\frac{\partial J_{*}}{(z-y)},\\
T_{*}(z)\,T_{*}(y) & \sim & \frac{2T_{*}}{(z-y)^{2}}+\frac{\partial T_{*}}{(z-y)}.
\end{eqnarray}
\end{subequations}Apart from $\beta$, all the other operators above
are primary fields and the central charge of the model vanishes.

As discussed in the main text, the ghost $\gamma$ should not have
any zeros on the worldsheet, otherwise the field redefinitions \eqref{eq:fieldredefinition}
are ill defined. This condition can be enforced by the parametrization\begin{subequations}
\begin{eqnarray}
\gamma & = & e^{\sigma},\\
\beta & = & \rho e^{-\sigma},
\end{eqnarray}
\end{subequations}and similarly for the hatted sector, such that
the holomorphic part of the action \eqref{eq:BU1action} and its associated
BRST charge are given by
\begin{eqnarray}
S_{*} & = & \int d^{2}z\{\bar{\Omega}\bar{\partial}\Omega+\rho\bar{\partial}\sigma\},\\
Q_{*} & = & \ointctrclockwise\,\Omega\rho.
\end{eqnarray}
Using the quartet argument, it is easy to show that the cohomology
of $Q_{*}$ contains only one element, the identity operator.

\section{A non-minimal formalism with fundamental $(b,c)$ ghosts\label{sec:NMBC}}

Because worldsheet reparametrization is a redundant symmetry of the
action \eqref{eq:newaction}, the gauge fixed model does not contain
the fundamental $(b,c)$ ghosts which are directly connected to the
definition of perturbative string theory as a sum over different worldsheet
topologies. In order to implement such structure in the pure spinor
formalism, Berkovits proposed the composite $b$ ghost in \cite{Berkovits:2001us},
which was later made super Poincar� covariant with the introduction
of the non-minimal formalism in \cite{Berkovits:2005bt}. In its simplest
form \cite{Oda:2005sd}, the pure spinor composite $b$ ghost can
be defined classically as
\begin{equation}
B\equiv\tfrac{\Lambda_{\alpha}}{4(\Lambda\lambda)}\Big(2\partial X^{m}(\gamma_{m}d)^{\alpha}+(\theta\gamma^{m}\partial\theta)(\gamma_{m}d)^{\alpha}+\tfrac{1}{2}(w\gamma_{mn}\lambda)(\gamma^{mn}\partial\theta)^{\alpha}+(w\lambda)\partial\theta^{\alpha}\Big),
\end{equation}
and satisfies $\{Q,B\}=T_{\tiny{PS}}$, where $Q$ is the BRST charge
\eqref{eq:PSBRSTcharge} and $T_{\tiny{PS}}$ denotes the holomorphic
component of the energy-momentum tensor associated to \eqref{eq:PSactiontraditional}
and can be expressed as
\begin{equation}
T_{\tiny{PS}}=-\tfrac{1}{2}\partial X^{m}\partial X_{m}-p_{\alpha}\partial\theta^{\alpha}-w_{\alpha}\partial\lambda^{\alpha}.
\end{equation}

When the master action \eqref{eq:PSMASTER} is extended to include
all the gauge and gauge-for-gauge symmetries described in subsection
\eqref{subsec:A-new-model}, it is possible to choose a gauge in which
the reparametrization ghosts survive in the form of non-minimal variables
such that the resulting BRST charge can be expressed as
\begin{equation}
Q=e^{-U}\Big(\ointctrclockwise\{\lambda^{\alpha}d_{\alpha}+b\phi\}\Big)e^{U},\label{eq:BerkovitsbcBRST}
\end{equation}
where $U$ is the generator of a similarity transformation given by
\begin{equation}
U=\ointctrclockwise\{-cB+\bar{\phi}c\partial c\}.
\end{equation}
Here, $b$, $c$, $\bar{\phi}$ and $\phi$ are the (fundamental)
Virasoro ghosts and ghost-for-ghosts with vanishing contribution to
the central charge of the action (-26+26=0). 

The BRST charge \eqref{eq:BerkovitsbcBRST} has the structure of a
generic coupling of the pure spinor superstring to topological two-dimensional
gravity, as analyzed in \cite{Hoogeveen:2007tu}, and was already
suggested in \cite{Berkovits:2014aia}, but it will be further explored
here. First note that it can be rewritten as
\begin{equation}
Q=\ointctrclockwise\{\lambda^{\alpha}d_{\alpha}+c(T_{\tiny{PS}}-2\bar{\phi}\partial\phi-\phi\partial\bar{\phi}-b\partial c)+b\phi\},
\end{equation}
such that the fundamental $b$ ghost satisfies
\begin{equation}
\{Q,b\}=T,
\end{equation}
where $T$ is the energy-momentum tensor of the non-minimal action,
\begin{equation}
T=T_{\tiny{PS}}-\bar{\phi}\partial\phi-\partial(\bar{\phi}\phi)-b\partial c-\partial(bc).
\end{equation}

Therefore, the similarity transformation in \eqref{eq:BerkovitsbcBRST}
helps to expose the familiar construction from gauge fixing worldsheet
diffeomorphisms. On the other hand, there does not seem to be any
relevant advantage in making such structure manifest. 

The whole machinery related to picture changing operators can be immediately
built. The bosonic ghosts $\phi$ and $\bar{\phi}$ can be \emph{fermionized}
as follows, 
\begin{eqnarray}
\phi & \cong & e^{\sigma}\eta,\\
\bar{\phi} & \cong & e^{-\sigma}\partial\xi,
\end{eqnarray}
such that the BRST cohomology of \eqref{eq:BerkovitsbcBRST} can be
written at picture $-1$ as
\begin{equation}
V^{(-1)}=e^{-U}\big(ce^{-\sigma}V\big)e^{U},
\end{equation}
where $V$ is an element of the pure spinor cohomology, with BRST
charge \eqref{eq:PSBRSTcharge}. Integrated vertex operators, $I$,
can be built with the action of the fundamental $b$ ghost on the
unintegrated vertex operators. In the picture $-1$, they can be built
as
\begin{eqnarray}
I^{(-1)} & \equiv & \ointctrclockwise\,b\cdot V^{(-1)},\nonumber \\
 & = & e^{-U}\Big(\ointctrclockwise\{e^{-\sigma}V+\ldots\}\Big)e^{U}.
\end{eqnarray}

Picture changing operators are also defined in a simple way. The picture
raising operator is given by
\begin{eqnarray}
Y & \equiv & \{Q,\xi\}\nonumber \\
 & = & c\partial\xi+(b-B)e^{\sigma},
\end{eqnarray}
while the picture lowering is given by
\begin{equation}
Z\equiv ce^{-\sigma},
\end{equation}
such that
\begin{equation}
\lim_{z\to y}Z(z)Y(y)=1.
\end{equation}
Note that because of the manifest spacetime supersymmetry, Neveu-Schwarz
and Ramond states are treated on equal footing and there are no half-integer
pictures.

At tree level, the integration measure for the reparametrization ghosts
and ghost-for-ghosts is simply
\begin{equation}
\left\langle c\partial c\partial^{2}ce^{-3\sigma}\right\rangle =1.
\end{equation}
$N$-point amplitudes can then be computed with $3$ vertices with
picture $-1$ to saturate the background charge ($\sigma$) and $(N-3)$
vertices with picture $0$.

In order to have a fully covariant formulation of this model, the
composite $B$ ghost introduced by Berkovits in \cite{Berkovits:2005bt}
can be used. Observe that for higher genus, $B$ will enter through
the picture changing operator, so it agrees with the usual pure spinor
prescription in which the $B$ ghost helps to saturate  the number
of fermionic fields ($d_{\alpha}$).

\section{The sectorized and ambitwistor pure spinor superstrings\label{sec:sectorized}}

The gauge \eqref{eq:conformalgauge} relies on the connection between
the twistor-like constraints and the generators of worldsheet reparametrization
in the first order formalism, represented through the identifications
displayed in \eqref{eq:H+-twistor}. As mentioned earlier, the gauge
$L^{\alpha}=\hat{L}^{\hat{\alpha}}=0$ would imply a degenerate worldsheet
metric. To see this, consider a different gauge fixing fermion of
the form
\begin{equation}
\Xi=\int d\tau d\sigma\{\bar{\Omega}A_{\tau}+\beta\chi_{\tau}-r^{\alpha}w_{\alpha}^{\sigma}+\eta^{\alpha}p_{\alpha}^{\sigma}-\pi_{\alpha}L^{\alpha}+\hat{\bar{\Omega}}\hat{A}_{\tau}-\hat{r}^{\hat{\alpha}}\hat{w}_{\hat{\alpha}}^{\sigma}+\hat{\beta}\hat{\chi}_{\tau}+\hat{\eta}^{\hat{\alpha}}\hat{p}_{\hat{\alpha}}^{\sigma}-\hat{\pi}_{\hat{\alpha}}\hat{L}^{\hat{\alpha}}\},
\end{equation}
thus implementing the singular gauge. The gauge fixed action can then
be shown to be
\begin{eqnarray}
\tilde{S} & = & \int d\tau d\sigma\{P_{m}\partial_{\tau}X^{m}+w_{\alpha}\partial_{\tau}\lambda^{\alpha}+p_{\alpha}\partial_{\tau}\xi^{\alpha}+\pi_{\alpha}\partial_{\tau}\theta^{\alpha}+r^{\alpha}\partial_{\tau}s_{\alpha}+\eta^{\alpha}\partial_{\tau}\epsilon_{\alpha}+A_{\sigma}\partial_{\tau}B\}\nonumber \\
 &  & +\int d\tau d\sigma\{\hat{w}_{\hat{\alpha}}\partial_{\tau}\hat{\lambda}^{\hat{\alpha}}+\hat{p}_{\hat{\alpha}}\partial_{\tau}\hat{\xi}^{\hat{\alpha}}+\hat{\pi}_{\hat{\alpha}}\partial_{\tau}\hat{\theta}^{\hat{\alpha}}+\hat{r}^{\hat{\alpha}}\partial_{\tau}\hat{s}_{\hat{\alpha}}+\hat{\eta}^{\hat{\alpha}}\partial_{\tau}\hat{\epsilon}_{\hat{\alpha}}+\hat{A}_{\sigma}\partial_{\tau}\hat{B}\}\nonumber \\
 &  & +\int d\tau d\sigma\{\Sigma\partial_{\tau}\chi_{\sigma}+\bar{\Omega}\partial_{\tau}\Omega+\beta\partial_{\tau}\gamma+\hat{\Sigma}\partial_{\tau}\hat{\chi}_{\sigma}+\hat{\bar{\Omega}}\partial_{\tau}\hat{\Omega}+\hat{\beta}\partial_{\tau}\hat{\gamma}\}.
\end{eqnarray}

The problem with this action is a residual gauge symmetry (aside from
the pure spinor symmetries) of the form $\delta\tau\to f(\tau)$,
a reparametrization of the worldsheet time coordinate. Therefore,
the singular gauge proposed in \cite{Berkovits:2015yra} does not
completely fix the gauge symmetries of the action \eqref{eq:newaction}.

There is, however, another interesting singular gauge. Consider now
the gauge fixing fermion
\begin{multline}
\Xi=\int d\tau d\sigma\{\bar{\Omega}A_{\tau}+\beta\chi_{\tau}-r^{\alpha}w_{\alpha}^{\sigma}+\eta^{\alpha}p_{\alpha}^{\sigma}+\hat{\bar{\Omega}}\hat{A}_{\tau}-\hat{r}^{\hat{\alpha}}\hat{w}_{\hat{\alpha}}^{\sigma}+\hat{\beta}\hat{\chi}_{\tau}+\hat{\eta}^{\hat{\alpha}}\hat{p}_{\hat{\alpha}}^{\sigma}\}\\
-\int d\tau d\sigma\{\pi_{\alpha}[L^{\alpha}-\tfrac{(\gamma^{m}\Lambda)^{\alpha}}{(\Lambda\lambda)}(P_{m}+\mathcal{T}\partial_{\sigma}X_{m})]+\hat{\pi}_{\hat{\alpha}}[\hat{L}^{\hat{\alpha}}+\tfrac{(\gamma^{m}\hat{\Lambda})^{\hat{\alpha}}}{(\hat{\Lambda}\hat{\lambda})}(P_{m}-\mathcal{T}\partial_{\sigma}X_{m})]\},
\end{multline}
which differs from \eqref{eq:GFF} by a sign in the last term. In
an analogy with the first order Polyakov action \eqref{eq:Pol1st},
this gauge can be interpreted as $e_{+}=-e_{-}=1$, \emph{cf}. \eqref{eq:H+-twistor}\footnote{I would like to thank Thales Azevedo for observing this is the gauge
choice leading to the bosonic sectorized model.}. In terms of the worldsheet metric, \eqref{eq:e+-tometric}, this
gauge is equivalent to $g_{\tau\tau}=0$. The gauge fixed action,
after the addition of a BRST expression, similarly to what is described
in subsection \eqref{subsec:Gauge-fixing}, can be written as
\begin{eqnarray}
\tilde{S} & = & \int d\tau d\sigma\{P_{m}\partial_{-}X^{m}+w_{\alpha}\partial_{-}\lambda^{\alpha}+p_{\alpha}\partial_{-}\xi^{\alpha}+A\partial_{-}B+\hat{w}_{\hat{\alpha}}\partial_{-}\hat{\lambda}^{\hat{\alpha}}+\hat{p}_{\hat{\alpha}}\partial_{-}\hat{\xi}^{\hat{\alpha}}+\hat{A}\partial_{-}\hat{B}\}\nonumber \\
 &  & +\int d\tau d\sigma\{\pi_{\alpha}\partial_{-}\theta^{\alpha}+r^{\alpha}\partial_{-}s_{\alpha}+\eta^{\alpha}\partial_{-}\epsilon_{\alpha}+\hat{\pi}_{\hat{\alpha}}\partial_{-}\hat{\theta}^{\hat{\alpha}}+\hat{r}^{\hat{\alpha}}\partial_{-}\hat{s}_{\hat{\alpha}}+\hat{\eta}^{\hat{\alpha}}\partial_{-}\hat{\epsilon}_{\hat{\alpha}}\}\nonumber \\
 &  & +\int d\tau d\sigma\{\Sigma\partial_{-}\chi+\bar{\Omega}\partial_{-}\Omega+\beta\partial_{-}\gamma+\hat{\Sigma}\partial_{-}\hat{\chi}+\hat{\bar{\Omega}}\partial_{-}\hat{\Omega}+\hat{\beta}\partial_{-}\hat{\gamma}\},
\end{eqnarray}
and all the worldsheet fields satisfy the equation of motion $\partial_{-}=0$,
constituting a chiral model. Furthermore, following the procedure
described in subsection \eqref{subsec:Fieldred}, the resulting action
is given by
\begin{equation}
S=\int d\tau d\sigma\{P_{m}\partial_{-}X^{m}+w_{\alpha}\partial_{-}\lambda^{\alpha}+p_{\alpha}\partial_{-}\theta^{\alpha}+\hat{w}_{\hat{\alpha}}\partial_{-}\hat{\lambda}^{\hat{\alpha}}+\hat{p}_{\hat{\alpha}}\partial_{-}\hat{\theta}^{\hat{\alpha}}\},\label{eq:Sectorizedaction}
\end{equation}
which corresponds to the sectorized string introduced in \cite{Jusinskas:2016qjd}.
The two chiral components of the BRST current can be cast as\begin{subequations}\label{eq:SectorizedBRST}
\begin{eqnarray}
J & = & \lambda^{\alpha}p_{\alpha}-\tfrac{1}{4\mathcal{T}}(\lambda\gamma_{m}\theta)(P^{m}+\tfrac{\mathcal{T}}{2}\partial_{+}X^{m})+\tfrac{1}{32\mathcal{T}}(\lambda\gamma_{m}\theta)(\theta\gamma^{m}\partial_{+}\theta),\\
\hat{J} & = & \hat{\lambda}^{\hat{\alpha}}\hat{p}_{\hat{\alpha}}-\tfrac{1}{4\mathcal{T}}(\hat{\lambda}\gamma_{m}\hat{\theta})(P^{m}-\tfrac{\mathcal{T}}{2}\partial_{+}X^{m})-\tfrac{1}{32\mathcal{T}}(\hat{\lambda}\gamma_{m}\hat{\theta})(\hat{\theta}\gamma^{m}\partial_{+}\hat{\theta}).
\end{eqnarray}
\end{subequations}

As a consequence of the singular gauge choice, the string tension
disappears from the action \eqref{eq:Sectorizedaction}, although
it is still present in the BRST current. Note also that the spacetime
spinors can be made dimensionless via a scale transformation of the
form
\begin{equation}
\begin{array}{rclcrcl}
\lambda^{\alpha} & \to & \mathcal{T}^{1/2}\lambda^{\alpha}, &  & \theta^{\alpha} & \to & \mathcal{T}^{1/2}\theta^{\alpha},\\
w_{\alpha} & \to & \mathcal{T}^{-1/2}w_{\alpha}, &  & p_{\alpha} & \to & \mathcal{T}^{-1/2}p_{\alpha},
\end{array}
\end{equation}
and similarly for the hatted fields. 

Now, the model has a well defined tensionless limit and the BRST currents
\eqref{eq:SectorizedBRST} are given by\begin{subequations}
\begin{eqnarray}
J & = & \lambda^{\alpha}p_{\alpha}-\tfrac{1}{4}(\lambda\gamma^{m}\theta)P_{m},\\
\hat{J} & = & \hat{\lambda}^{\hat{\alpha}}\hat{p}_{\hat{\alpha}}-\tfrac{1}{4}(\hat{\lambda}\gamma^{m}\hat{\theta})P_{m},
\end{eqnarray}
\end{subequations}which correspond to the ambitwistor pure spinor
superstring \cite{Berkovits:2013xba}, in agreement with the results
of \cite{Azevedo:2017yjy}.

The heterotic sectorized or ambitwistor strings in the pure spinor
formalism are obtained in a similar way. Worldsheet reparametrization
and only one of the twistor-like constraints has to be taken into
account in building the master action \eqref{eq:PSMASTER}. Apart
from that, the gauge fixing procedure should be very similar.

\end{document}